\documentclass[a4paper,twoside,10pt]{article}
\usepackage{graphicx,publaob}

\begin{document}

\title{Neutrino Oscillations and the Early 
Universe}

\authors{D. P. Kirilova}

\address{Institute of Astronomy,  BAS, Sofia~\footnote{Regular
Associate of Abdus Salam ICTP}, 
 and  Physique Theorique, ULB, Bruxelles}
\markboth{Neutrino Oscillations and the Early Universe}{D. P. Kirilova}

\abstract{The observational and theoretical status of neutrino oscillations 
in connection with solar and atmospheric neutrino anomalies is presented 
in brief. 
The effect of neutrino oscillations on the early Universe evolution 
is discussed in detail. A short review is given of the standard Big Bang 
Nucleosynthesis and the influence of resonant and nonresonant neutrino 
oscillations on active neutrinos and on primordial synthesis  of 
He-4. 
     BBN cosmological constraints on neutrino oscillation 
parameters   are discussed.}

\section{INTRODUCTION}
 
Neutrino - a neutral weakly interacting particle, is of extreme interest for 
Physics and Astrophysics. It is a key to the investigation of the weak 
interactions and the physics beyond the standard electroweak model. 
On the other hand, being a very weakly interacting particle, and hence having 
a uniquely great  
penetrating capability, neutrinos carry precious information  for the 
astrophysical 
processes in the most dense regions of the star cores and from the very early 
stages of the Universe evolution.
Therefore, revealing neutrino characteristics is of great importance.

The contemporary particle physics theory neither requires nor forbids 
a nonzero neutrino mass. In the standard model of particle physics 
neutrinos are assumed massless. 
In the more general case of non-zero neutrino masses, the weak neutrino 
eigenstates 
may be a linear combination of the mass eigenstates, which means that 
transitions between neutrinos with different types (flavours), the so called 
{\it neutrino oscillations} are possible. 
Neutrino oscillations and their role in resolving the solar neutrino puzzle 
were first proposed by B. Pontecorvo (see Pontecorvo, 1958) and after   
45 years 
they continue to be the theme of leading experimental and theoretical 
research. 

 In recent years  positive indications for neutrino oscillations were 
obtained at the greatest neutrino
experiments (evidence for solar neutrino  oscillations: Homestake,
Kamiokande,
SuperKamioKa, Gallex, SAGE, SNO; evidence for atmospheric
neutrino oscillations:
Super-KamioKa,
Macro, Soudan 2, IMB; evidence for neutrino oscillations at terrestrial
experiments: LSND, KamLAND, K2K)
(see refs. Gonzalez-Garcia \& Nir, 2003; Smirnov, 2003; Giunti  \&
Laveder, 2003). Each of
these neutrino
anomalies, namely the
solar neutrino problem, atmospheric neutrino anomaly and the
positive results of terrestrial LSND and KamLAND experiments may
be resolved by the phenomenon of neutrino oscillations.
These results have a great resonance as far as any experimental evidence
for neutrino masses or mixing is a signal of
new physics (NP)  - physics beyond the standard model of electroweak
interactions.

 On the other hand,  neutrino oscillations affect early Universe 
evolution
by affecting expansion rate,  neutrino densities and neutrino energy
spectrum,  neutrino - antineutrino asymmetry, 
thus influencing the neutrino involved processes, as for example 
cosmological
nucleosynthesis, structure formation, etc.
Cosmological nucleosynthesis, traditionally called
Big Bang Nucleosynthesis (BBN)  explains very successfully
the data on the primordial abundances of the light elements D,
He-3, He-4 and  Li-7, and
is traditionally used as a probe of the
conditions of the early Universe, due to the high accuracy
of the theoretically predicted abundances of light elements
  and to the good
accuracy of their primordial values inferred from observations.
 Hence, BBN  is a
 powerful probe for NP, like  neutrino
oscillations.

From BBN considerations most stringent constraints
on
neutrino oscillations parameters are obtained. In particular,
LMA and LOW active-sterile solar oscillation solutions and
atmospheric active-sterile solutions were excluded many years
before the global analysis of experimental neutrino data pointed
to the preference of
flavour oscillations for solving these neutrino anomalies.
 
In the following we will present a brief introductory review of the 
solar and atmospheric neutrino anomalies and then discuss in more detail
the role of neutrino oscillations in the early Universe and the
cosmological constraints on oscillation parameters, following from
BBN.

\subsection{Neutrino oscillations} 

The basic idea of {\it neutrino oscillations} is
that left-handed mass eigenstates $\nu_i$ are distinct from the
left-handed flavour eigenstates $\nu_f$:
$$
\nu_i=U_{if}~\nu_f~~~~~ (f=e,\mu, \tau).
$$
Then in the simple two-neutrino oscillation case in vacuum, the 
probability to find
after a time interval $t$  
a given neutrino type in an initially homogeneous neutrino beam of the
same type is:
$P_{ff}=1-\sin^22\vartheta \sin^2 (\delta m^2 t/4 E)$, 
 where $\delta m^2$ - the neutrino mass difference  and 
$\vartheta$ - the oscillations mixing angle are the oscillation
parameters, $E$ is the neutrino energy. 
I.e. the flavour composition changes with time. 

 The medium distinguishes between different neutrino types due
to 
different interactions (Wolfenstein, 1978; Mikheyev \& Smirnov, 1985). 
This leads to different average potentials $V_f$  
for different neutrino types.

$$
V_f=Q \pm L
$$ 

 \noindent where $f=e, \mu, \tau$, $Q=-bET^4/(\delta m^2M_W^2)$,
$L=-aET^3L^{\alpha}/(\delta m^2)$,
$L^{\alpha}$ is given through the fermion asymmetries of the plasma,
$a$ and $b$ are positive
constants different for the different neutrino types,
$-L$ corresponds to the neutrino and $+L$ to
the antineutrino case. The sterile neutrino does not feel  
 the medium, hence $V_s=0$. 

The effects of the medium can be hidden in  $\delta m^2$ and 
$\vartheta$. Namely, 
the matter mixing angle in the adiabatic case is 
expressed through the vacuum oscillation parameters
and the characteristics of the medium, like its density and temperature.
For the early Universe the following relation holds:
$$
\sin^2\vartheta_m=
\sin^2\vartheta/[\sin^2\vartheta+(Q \mp L-\cos2\vartheta)^2],
$$
Although  in general the medium suppresses
oscillations by  decreasing their amplitude,
there also exists  a possibility of enhanced 
oscillation transfer in case a resonant condition between the
parameters of the medium
and the oscillation parameters holds:
$$
Q \mp L=\cos2\vartheta.
$$
\noindent Then the mixing in matter becomes maximal, independently of the
value of the vacuum mixing angle, i.e. resonant transfer takes place.

At high temperature of the early Universe for a lepton asymmetry of the 
order of the baryon one, $Q > L$. So for $\delta m^2 <0$ resonance is 
possible both for neutrino and antineutrino. At low $T$, however, 
 $L > Q$, and as can be seen from the resonant condition, if $\delta m^2 
>0$ a resonance in the neutrino ensemble can take place, while for $\delta 
m^2 <0$ - the resonance is possible only for  the antineutrinos. 

 Both the nonresonant and resonant
 oscillation cases are interesting from a cosmological
point of 
view and  from the viewpoint of the 
discussed neutrino anomalies.

\section{NEUTRINO ANOMALIES AND NEUTRINO OSCILLATION EXPERIMENTS}

\subsection{SOLAR NEUTRINO DEFICIT}
 
According to the contemporary astrophysical understanding the Sun is 
a Main Sequence star at the stage of hydrogen burning. It produces an intense 
flux of electron neutrinos as a result of its nuclear reactions generating the 
solar energy.
Due to its weak interaction with matter, the solar neutrino reaching the Earth 
comes from the very deep solar core and carries valuable information 
about stellar structure and its evolution. Hence, the detection of the  
neutrino from the Sun has been recognized as a task of great importance 
as early as the 50ies - when Davis started a radiochemical experiment, 
aiming to detect neutrinos from the Sun,  in the golden mine of Homestake.
The solar neutrinos also present the unique possibility for investigation 
of the neutrino properties like neutrino mass and mixing, because the Sun is 
at a  very large distance from the Earth, and also because the solar density 
varies strongly from the center to the surface and thus offers interesting 
conditions for the penetration of neutrino through layers with different 
density and thickness.   

Since the first attempts to measure solar neutrinos,  there has been  
different types of solar neutrino experiments, 
using Cl, Ga and H$_2$O and D$_2$O as targets for measuring 
electron neutrino from the
Sun.
The detected fluxes of solar neutrinos at these solar experiments 
(using different detection methods
\footnote{There exist radiochemical experiments like GALLEX, SAGE and 
Homestake and electron experiments like Kamiokande 
and SuperKamioka.}
 and sensitive to  different energy ranges) 
are in qualitative agreement with 
the assumption that Sun burns due to nuclear reactions in its core. 
  However, {\it all 
the data of the 
solar neutrino experiments point to a considerably lower neutrino flux 
than the expected one in the standard Solar Model. 
Furthermore, the suppression is different in various experiments, 
sensitive to different energy range.} This problem is called 
{\it solar neutrino anomaly}. 

Despite the continuous 
improvements of the Solar Model and the 
predicted neutrino flux in the last 40 years, the discrepancies between the 
observations and the predictions of the model persist.
Depending on the energy the measured fluxes  
consist   0.3 to 0.6 of the predicted values.
 The recent measurement of neutral currents and charged 
currents 
fluxes at SNO experiment (Ahmad et al., 2002) provide $\sim 5 \sigma$ 
signal for neutrino 
flavour 
transitions that is not strongly dependent on the Solar Model.  
 Hence, 
there remain  less doubts about the Solar Model prediction
capability.

Thus, in case we exclude the possibility that most of the leading solar 
neutrino experiments are wrong, the experimental data points more and 
more convincingly to the necessity of new neutrino physics. 

Neutrino oscillations are  capable to explain the observational data and 
its discrepancies with the predictions of the Solar Model:  
The electron neutrino, produced in the solar core,
 undergoes 
transformations into other flavours  
while penetrating   through the Sun and  the cosmic space till the  
terrestrial detectors of electron neutrinos.  
Hence, the registered electron neutrino flux is  reduced in 
comparison with the flux produced in the Sun core.~\footnote {Except the 
flux measured by SNO, which detects  
 all flavour neutrinos, not only electron 
ones.}  

There  existed different types of solar neutrino oscillations 
solutions - 
Small Mixing Angle  (SMA) and Large Mixing Angle (LMA),  depending on the 
mixing angles at around $\delta m^2\sim10^{-5}$ eV$^2$ and 
LOW and vacuum oscillation solutions corresponding to very small 
$\delta m^2\le10^{-7}$ eV$^2$ mass differences and maximum mixing.  
 The present solar neutrino data  
  definitely prefers flavour oscillation solutions to  
active-sterile ones (as was pointed first from cosmology considerations - 
see  section 4), and  in the light of the recent results of the 
terrestrial experiment KamLAND,
the LMA solution   is split into two sub-regions, and is the 
chosen 
one.  The best fit point values of $\delta m^2$ are  
$(7.3 \pm 0.8).10^{-5}$ eV$^2$  and   $\sin^22\theta\sim0.315 \pm 
0.035$. For 
more details see ( Bahcall et al., 2003, Balantekin  \& Yuksel, 2003,  
Holanda \& Smirnov, 2003; Fogli et al., 2003; Maltoni et 
al., 2003). 

\subsection{ATMOSPHERIC NEUTRINO ANOMALY}

A continuous isotropic flux of cosmic rays, consisting of protons and heavy 
nuclei, is bombarding the Earth's atmosphere. As a result of its interactions 
with the atmospheric particles pions and kaons are produced, which decay 
and produce  muon and electron neutrinos  with 
a wide energy range. The theoretical prediction for the ratio of the muon to 
the electron flux is $r \sim 
(\nu_{\mu}+\bar{\nu_{\mu}})/(\nu_e+\bar{\nu_e})=2$ 
for energies less than $1$ 
GeV. 
Besides, identical up-coming and down-coming fluxes are expected due to the 
isotropy of the cosmic rays flux and due to the 
spherical symmetry of the Earth's atmosphere. Any deviation from these 
predictions is an indication for new neutrino physics.  

The underground neutrino experiments SuperKamioka, Soudan 2 and Macro, as 
well as the earlier experiments IMB and Kamiokande, have {\it measured 
ratio $r$ 
considerably lower than the expected one} (see for example Fukuda, 1998). 
This discrepancy, the so called 
{\it atmospheric neutrino anomaly} is known already for more than 10 years. 
Besides a dependence of the muon neutrino deficit on the  
 zenith angle and distorsion of the energy spectrum  is observed.
 
The experimental data can be explained in terms of neutrino oscillations,  
namely by the transition of the muon neutrino into another type. 
The latest data analysis indicates the $\nu_{\mu}\leftrightarrow\nu_{\tau}$
channel as the dominant one. The oscillations into sterile neutrino are 
disfavoured, because of the absence of suppression of  
oscillations by the medium, expected in the sterile case at high energies.  
The best fit oscillation parameters for the available data are 
nearly maximal mixing 
 and $\delta m^2 \sim (2.6 \pm 0.4) \times10^{-3}$ 
eV$^2$. 
For more detail see (Guinti, 2003 and references 
therein). 

\subsection{LABORATORY EXPERIMENTS}

Besides these two astrophysical indications for neutrino oscillations
 and non-zero neutrino mass, there exist also 
laboratory experiments, the so-called terrestrial experiments LSND 
(Aguilar et al., 2001), K2K (Ahn et al., 2003) and 
KamLAND (Eguchi et al., 2003),  
which data have given an indication for oscillations, too. 

{\it The short baseline Los Alamos Liquid Scintillation Neutrino Detector} 
(LSND) 
 experiment  has registered 
appearance of electron antineutrino in a flux of muon antineutrino. 
This anomaly might be interpreted as $\nu_{\mu}\leftrightarrow\nu_e$ 
oscillations with $\delta m^2=O(1$~ eV$^2)$
 and $\sin^22\theta=O(0.003)$. 

In case LSND result is confirmed\footnote{The LSND result was not confirmed 
by KARMEN. It  will be tested in future by the ongoing MiniBoom 
experiment at Fermilab (Bazarko, 2002).} 
  an addition of a light singlet 
neutrino (sterile neutrino $\nu_s$) is required, because three different 
mass differences, needed for the explanation of the solar, atmospheric 
and LSND anomaly require 4 different neutrino masses. 
 This simple extension 
already has difficulties, because oscillations into purely sterile 
neutrinos do not fit neither the atmospheric nor solar neutrino data. 
Both 2+2 and 3+1 oscillation schemes have problems. Active-sterile 
oscillations are strongly restricted by 
cosmological considerations as well (for review on 2 oscillation 
constraints see Kirilova \& Chizhov, 2001; and for 4 neutrino oscillation
 schemes see   
(Bilenky et al., 1998; Di Bari, 2002; Dolgov \& 
Villante, 2003). 

{\it K2K, a long baseline neutrino  experiment},  has probed the 
$\delta m^2$ region explored by atmospheric neutrinos. It has measured 
muon neutrino deficit in a beam coming from KEK 
to Kamiokande. The distance is $250$ km  and $E\sim 1.3$ GeV. $56$ events 
were observed instead of the expected $80 \pm 6$ in case without 
oscillations. A hint of energy spectrum distortion is also indicated by 
the 
analysis. 

The results are consistent with SuperKamioka atmospheric data and confirm 
atmospheric neutrino solution. 

 {\it KamLAND (Kamioka Liquid Scintilator Anti-Neutrino Detector)}  
experiment
explored with reactor neutrinos the region of oscillation parameters 
relevant for the solar neutrinos. It  
 has measured electron antineutrino deficit in the flux of antineutrinos 
coming from reactors at $\sim 180$ km distance. 
In the context of two-flavour neutrino oscillations KamLAND results 
single out   
  LMA  solution,  as the  oscillation solution to the 
solar neutrino problem.   The allowed previously LMA 
region is further reduced by its results (see Eguchi et al., 2003).   
So, the KamLAND result appears to confirm in a totally independent and
completely 
terrestrial way  that solar neutrino deficit is 
indeed due to neutrino oscillations, which was suspected in  many solar 
neutrino experiments over the last 40 years. 
 
KamLAND reactor and KEK accelerator experiments strongly contributed to 
the reduction of the allowed range of mass differences for solving the 
solar and atmospheric neutrino anomalies. Their results mark the 
beggining of the precision epoch in determinations of neutrino 
characteristics. 

The neutrino experiments results confirm non-zero neutrino mass and mixing. 
Non-zero neutrino mass is also cosmologically welcome, as it may 
play the role of 
the hot dark matter component essential for the successful structure
 formation (see the next section).  
The standard model of particle physics (SM) 
$SU_c(3)\times SU_W(2)\times U_Y(1)$  does not predict  non-zero neutrino 
mass and mixing. To explain the smallness of the neutrino mass differences  
 new physics  beyond SM is required. Hence, neutrino data gathered at 
neutrino 
experiments 
and the 
cosmological considerations concerning neutrino mass and mixing, discussed 
below, point the way towards this NP - hopefully the true unified theory of 
elementary particles.  

\section{NEUTRINOS IN THE EARLY UNIVERSE}

After the photons of the microwave background radiation, neutrinos are the 
most abundant particles in the Universe.~\footnote{Their present day 
number density 
is $n_{\nu}=n_{\bar{\nu}}\sim 56$ cm$^{-3}$ for each neutrino flavor.} 
Hence, 
in case they have non-zero 
mass they may contribute considerably to the total energy density of the 
Universe. From the requirement that the neutrino density should not exceed 
 the  matter density, and asuming that the Universe is older than the Earth, 
an upper bound  on the 
neutrino mass was 
derived -- "Gerstein-Zeldovich" limit (see Gerstein and 
Zeldovich, 1966). The contemporary version of it reads:
$$
 \Sigma m_{\nu_f} \le 94  eV \Omega_m h^2=15 eV, 
$$
where 
$\Omega_m$ is the matter density in terms of the critical density and $h$ 
is 
the dimensionless Hubble parameter. In deriving this limit 
$\Omega_m<0.3$ and $h=0.7$ is assumed 
according to 
contemporary astronomical data.  
 Hence, the neutrino
mass of any neutrino flavour should be less than about $5$
eV.

Much stronger limits on the neutrino mass may be obtained 
accounting for the considerable role of neutrinos in other important 
cosmological processes, like the primordial nucleosynthesis and the formation 
of large scale structure of the Universe, the formation of the cosmic 
microwave background radiation, etc. 

\subsection{Large Scale Structure and Neutrino}

The mass of the visible (radiating) matter in the Universe is at most $0.01$ 
of the total mass 
deduced from its gravitational effect. The 
remaining  $0.99$ 
consists of the  so-called {\it Dark Matter} of the Universe (DM). Only 
a negligible portion of this DM may be in the form of invisible baryons, 
i.e. $0.04$ of 
the total mass, i.e. DM is mainly  
non-baryonic. Massive neutrinos with a mass of a few eV could  naturally 
be the  
candidates for 
DM component in clusters of galaxies (Cowsik \& McClleland, 1972). 

On the other hand, according to the accepted contemporary theory,   
structures in the Universe are a result of  
gravitational instabilities of overdensity perturbations. 
These perturbations result from the initial microscopical 
perturbations generated at the inflationary stage, which have been inflated 
during the exponential expansion.
Neutrinos cluster more efficiently in larger potential wells.  
In case they play the role of DM mass-to-light ratio should increase 
with scale. That behavior was not confirmed by observations.
 Moreover,
 to be in agreement with the sizes of the  structures
 observed today,
it was found necessary to speed up the growth of the perturbations, which is
naturally achieved by the presence of non-relativistic DM at the epoch of
perturbations growth, which should be the main DM component. 
So, massive neutrinos are not the main DM component in galaxies and 
clusters. 

Still, the precise 
analysis of the recent microwave background
anisotropy data and  structures data at large red-shifts points  
to the necessity of 
some admixture of hot DM
 which can be 
naturally provided by  
light neutrinos with $\Sigma m_{\nu_f} < 2.2$ eV or $m_{\nu_f}<0.73$ 
eV (see for example Fukugita et al., 2000; Elgaroy et al., 
2002).
Recent WMAP measurements (see Spergel et al., 2003)  together with the 
LSS analysis improved  the limit: 
$\Sigma m_{\nu_f} < 0.69$ eV and hence, $m_{\nu_f}<0.23$ eV.
See also recent analysis of the CMB, LSS and X-ray galaxy cluster data 
(Allen et al., 2003), which give for the preferred non-zero neutrino mass  
 the  bounds: $\Sigma m_{\nu_f} \sim 0.64$ eV or $m_{\nu_f} \sim 
0.21$  eV per neutrino. 

It is remarkable that such mass value is in accordance with the picture of  
 oscillation models, which predict oscillations between nearly 
degenerate 
neutrinos, with negligibly different masses (in case of solar neutrino 
anomaly with mass difference $\sim10^{-5}$ 
eV$²$ 
and in case of the atmospheric anomaly -  $\sim0.001$ eV$²$). Each of these 
degenerate  neutrino types, in case 
they have masses of the order of $\sim 0.2$ eV, can successfully play the 
role of the hot DM. 

For comparison the laboratory bound on electron neutrino mass from Tritium 
$\beta$-decay  experiments is $m_{\nu_f} < 2.2$ eV. Given the small mass 
differences pointed from solar and atmospheric neutrino data, this bound 
applies to each neutrino eigenstate, i.e.  $\Sigma m_{\nu_f} < 6.6$ eV 
(Barger, 1998). So, the cosmological bounds on neutrino masses at present 
are more restrictive than the laboratory constraints.  
For 
recent 
review on laboratory measurements of neutrino masses 
and also their cosmological and astrophysical constraints see  
(Bilenky et al.,
2003; Sarkar, 2003; Dolgov, 2002; Raffelt, 2002).

\subsection{BBN and neutrino}

One of the most exciting events in the early Universe is the primordial 
nucleosynthesis of the light elements. 
The idea for the production of elements through nuclear reactions in the 
hot plasma during the early stage of the Universe evolution belongs to 
George Gamov and was proposed and developed in the 1930s and 1940s (Gamow, 
1935, 1942, 1946). 
In the following 70 years this idea 
has grown to an elegant and famous theory - theory of the cosmological 
nucleosynthesis (Big Bang Nucleosynthesis), explaining successfully the data 
on the abundances of  D, He-3, He-4 and Li-7
 (see for example Esposito et al., 2000, and references therein). 

Recently BBN theory was improved: nuclear data have been reanalysed, 
all nuclear reactions rates involving $\sim 100$ processes were updated, 
new processes were included, estimates of the weak rates were improved. 
Also statistical uncertainty of observational determination of the 
light elements was improved. And finally, thanks to the 
presice determination  
of the baryon density in CMB anisotropy measurements, it  
 was used as an input in BBN.   
The present status of BBN after the recent measurements of CMB 
anisotropies by WMAP experiment is presented in detail in (Cuoco et al., 
2003; 
Cyburt et al., 2003: Coc et al., 2003, Steigman, 2003).    

Based on the excellent agreement between CMB, BBN predictions and the 
observational  data, today we believe that we know to a great precision 
the physical processes typical for the BBN epoch. Hence, BBN is a  
most powerful probe for new physics, like the physics predicting neutrino 
oscillations and non-zero neutrino mass.
\ \\ 
 
According to the Standard BBN  model (SBBN) the cosmological nucleosynthesis 
proceeds when the temperature of the 
plasma falls down to 1 MeV, when  the weak processes, governing the 
neutron-proton transitions become comparable with the expansion rate.
 As a result the neutron-to-proton ratio freezes out at temperature  
below  $0.7$ MeV. This ratio 
enters in the following rapid nuclear reactions leading to the synthesis of D
and the rest light elements formed in the first hundred seconds from the 
Big Bang.

Only at  $T<80$ keV, a temperature well below the $D$ binding energy 
$\sim 2.2$ MeV, 
 the building of complex nuclei becomes possible, 
the  first step being: $n + p = D + \gamma$.  At 
higher $T$ deuterons were quickly photo-disociated because of the large 
photon-to-baryon ratio $\eta^{-1}$.  

So, actually in SBBN, $\eta$ is the only parameter.
Most sensitive to  $\eta$ among the light elements is D, therefore it was 
considered till recently the best baryometer. 
The observational $\eta$ values
 preferred today are namely,  the one obtained on the basis of
measurements of D  in high redshift
QSO Absorption Line Systems  of
 (Kirkman et al., 2003), namely $\eta \sim (6.1 \pm 0.5)\times 10^{-10}$ and
  $\eta$ obtained from the
CMB anisotropy measurements $\eta = (6.1 \pm 0.25) \times 10^{-10}$ 
(Spergel et al.,
2003). Present day CMB measurements of  $\eta$ are considered tighter
than the BBN one, so  $\eta_{CMB}$ is used as an input for BBN
calculations.
 
According to the standard BBN during the early hot and dense
epoch of  the Universe
   only 
D, $^3$He, $^4$He, $^7$Li were synthesized in considerable ammounts.
 $^4$He is with the highest binding energy among the light nuclides,
hence $D$ and $^3$He were rapidly burned into it.
 $^4$He is the most abundantly produced. 
The production of heavier elements was hindered by the rapid 
decrease of the Universe density with the cooling of the Universe, 
growing Coulomb barriers and the absence of a stable mass 5 nuclide.  
The  latter were formed much later in stars. 

He-3 and Li-7 have a complex post BBN evolution.
They are both created and distroyed in stars, hence are unreliable as a
cosmological probes.
Besides both their theoretically calculated  and observational  values
suffer from large uncertainties.
D although having a clear post BBN chemical evolution, namely it is
believed to have been only distroyed after BBN, as far
as it has the
lowest binding energy of the light nuclides,  still has large
theoretical and observational uncertainties - up to $10\%$:
$D/H=(2.6 \pm 0.4) 10^{-5}$ (Kirkman et al., 2003). Besides, there is
a significant dispersion among the derived D abundances at low
metalicity
Z, a fact suggesting either the existence of systematic errors or
a revision of our concepts about D post BBN evolution.
 
On the contrary,  $Y_p$, predicted by BBN, is
calculated
with  great precision (see Lopez and Turner, 1999;
Esposito et al., 2000, Cyburt et al., 2003, Cuoco et al.,
2003, Coc et al., 2003).
The theoretical uncertainty
is less than  0.1\% ($|\delta Y_p| < 0.0002$) within a wide range of
values of the baryon-to-photon ratio $\eta$.
The predicted He-4 value
is in in relatively good accordance with the
observational data for He-4 and is consistent with other light
elements abundances.    The contemporary helium values, inferred from
astrophysical observational data, are 0.238--0.245
( Olive et al., 1997; Izotov \& Thuan, 1998).  And 
although there exist some tension between the two different
measurements giving different He values, 
and also  between the observed He values and 
the predicted ones, using $\eta$ indicated either by D measurements or by 
CMB, 
taking the central Helium value of the 2 measurements $0.238$ and
 assuming  the systematic error of
$0.005$ reestablishes the agreement both between different helium 
measurements 
and
also between observed and the predicted He-4 values. I.e. there is an
accordance at $2 \sigma$ level and the uncertainty is only around $2\%$.

 Measurements of primordial helium
from CMB data are possible.
 Hopefully future Planck CMB
measurements will be capable
to determine the helium mass fraction within $\delta Y \sim 0.01$  in
a completely independent way (see Trotta and Hansen, 2003).
 Primordial helium value
is also in a good accordance
 with the initial helium content, necessary for the successful star
evolution modelling (Bono et al., 2002; Cassisi et al., 2003).

 In conclusion,  $^4$He is the most abundantly produced ($\sim 24$\% by 
mass), most precisely measured ($\sim 2$\%  uncertainty) and
most precisely calculated element  ($\sim 0.1$\%  uncertainty) . This 
fact 
and  its relatively simple chemical
evolution
make it  the preferred element
 for  probing non-standard physics.  Particularly, 
for the  analysis of
the oscillations effect on BBN, He-4 is 
the  traditionally used element, as well.
Therefore, we will discuss it in more detail below. 

According to the standard cosmological nucleosynthesis (SBBN)  
 He-4 primordial yield  essentially 
depends  on the freezing  
of the reactions interconverting neutrons and protons:
$$
\nu_e+n\leftrightarrow p+e^-,~~ e^++n\leftrightarrow
p+\tilde{\nu_e},
$$
 which  maintain the equilibrium 
of nucleons at high temperature ($T>1 MeV$) $n/p\sim exp(-\Delta m/T)$, 
where $\Delta m=m_n-m_p$, $T$ is the temperature 
$T=T_{\gamma}=T_e=T_{\nu}$ prior to electro-positron annihilation. 
For the radiation dominated epoch $\rho \sim \rho_{\gamma}+ \rho_{\nu}+ 
\rho_e=g_{eff} T^4$.                       
The $n/p$ freeze-out occurs when in the process of expansion 
the rates of these weak processes
$\Gamma_w \sim G_F^2 E_{\nu}^2 N_{\nu}$
become 
 comparable and less than the expansion rate  
$H(t)=8 \pi G \rho/3 \sim\sqrt{g_{eff}}~T^2$.
As far as the temperature of freezing is $T_f \sim 
g_{eff}^{1/6}$ the neutron-to proton frozen ratio $(n/p)_f$ is sensitive 
to $g_{eff}$. 

Further evolution of the neutron-to-proton ratio is due to the neutron 
decays that proceed until the
effective synthesis of D begins.
Almost all available neutrons are sucked
into He-4.
So, the primordially
produced mass fraction of He-4, to a good
approximation, is 
$$
Y_p(\mbox{\rm He-4})\sim2(n/p)_f/(1+n/p)_f \exp(-t/\tau_n). 
$$
 Hence, the produced He-4 is a strong function of the
 effective number of relativistic degrees of freedom at BBN epoch,
$g_{eff}=10.75 +7/4 \delta N_s$, neutron mean lifetime $\tau_{n}$, 
which parametrizes
the weak interactions strength.  He-4 depends weakly on  the nucleon-to 
photon ratio $\eta$.
It  depends also on the electron neutrino 
  spectrum  and  on the neutrino-antineutrino
asymmetry, which enter through  
$\Gamma_w$.
In the standard BBN model three neutrino
flavors ($\delta N_{\nu}=0$), zero lepton asymmetry and equilibrium   
neutrino
number densities and spectrum distribution are postulated:
$$
 n_{\nu_e}(E)=(1+\exp(E/T))^{-1}.
$$

Due to its strong dependence on $g_{eff}$ $He-4$ abundance is used to 
constrain the 
number of the relativistic during BBN particles (Shvartsman, 1969), 
usually parametrized by   $\delta N_{\nu}$.  
For contemporary discussion of BBN constraints on  $\delta N_{\nu}$ 
see for example  (Lisi et al., 1999). The present BBN upper bounds  on  $\delta 
N_{\nu}$  depending on the concrete 
analysis vary in the range $\delta N_{\nu}<0.1$ -  $\delta N_{\nu}<0.7$
 (Barger et al., 2003, Cyburt et al, 2003; Cuoco et al., 2003). 
The  value we consider reliable enough  for putting cosmological 
constraints on new physics parameters, corresponding to $\sim 3\%$  
overproduction of He-4  is  $\delta N_{\nu} < 0.64$.

BBN  constraint is in  agreement with the constraints based 
on LSS and WMAP data (Crotty et al., 2003). 

The dependence of the primordial abundances  on the density 
and on the nucleon kinetics was used also for constraining massive 
stable neutrinos and decaying massive neutrinos (Terasawa \& 
Sato, 1987; Dolgov \& Kirilova, 1988; Gyuk  \& Turner 1994). The 
contemporary status of these 
constraints is presented in (Dolgov et al., 1999).

For more details on  neutrino role
in cosmology see (Dolgov, 2002, 2003) and the  references
therein. In the following we will concentrate mainly on BBN and neutrino
oscillations.
\ \\

\section{BBN WITH NEUTRINO OSCILLATIONS}

The influence of neutrino oscillations depends on the type of oscillations:
oscillation channels, resonant transitions, the degree
of equilibrium of oscillating neutrinos (see the review of  
Kirilova \& Chizhov, 2001 and the references therein).
Flavour neutrino oscillations effect BBN
negligibly in case 
 different flavour neutrinos  are in 
thermal equilibrium and with vanishing chemical potentials (Dolgov, 81).
However,  active-sterile oscillations may 
have considerable influence  because they  
  effect both expansion rate through
exciting additional neutrino types,
and the weak interactions rate due to  shifting neutrino densities and
energy spectrum  from BBN equilibrium values.

On the other hand although  solar and atmospheric neutrino anomalies 
can be explained without a sterile neutrino, and definitely do not allow 
active-sterile neutrino oscillations
 as a dominant channel, some subdominant admixture of steriles is not only 
allowed, but also desirable (Holanda \& Smirnov, 2003).
Hence, it is interesting to discuss the cosmological effects of 
active-sterile neutrino oscillations, and to provide cosmological 
limits to oscillation parameters, which can be helpful in constraining 
the possibilities for $\nu_s$ admixture.  

 The presence of
 neutrino oscillations invalidates  BBN assumptions
about
three neutrino flavours,
zero lepton asymmetry,
equilibrium neutrino  energy
distribution, thus  directly
influencing the kinetics of nucleons during the weak freeze-out.

\subsection{Neutrino Oscillations Effects}

{\it Qualitatively, 
neutrino oscillations effects considerably influencing  the neutrino 
involved processes
in 
the Universe are} 

{\it (a)~Excitation of additional degrees of freedom}: 
 This  leads to faster Universe expansion  $H(t)\sim g^{1/2}_{eff}$,
earlier $n/p$-freezing,  $T_f\sim (g_{eff})^{1/6}$,
at times when
neutrons were more abundant (Dolgov, 1981)
$$
n/p\sim \exp(-(m_n-m_p)/T_f)
$$
This effect gives up to $5\%$ $^4\!$He overproduction
(in case  one additional neutrino type is brought into
equilibrium by
oscillations,  $\delta N_s=1$).

{\it (b)~Distortion of the neutrino spectrum}: 
The  effect of oscillations may be  much stronger than $\delta N_s=1$ in
case of
oscillations effective after $\nu$ decoupling,  proceeding between
partially populated
sterile neutrino state $0\le\delta N_s<1$ and
electron neutrino (Kirilova 1988; Kirilova and Chizhov, 1996; Kirilova,
2002). The non-equilibrium initial condition, 
for most of the oscillations parameters of the model, leads to
considerable and continuous deviations from the equilibrium $\nu_e$
 spectrum
(spectrum distortion).

A study of the momentum dependent kinetic equations for
oscillating neutrinos  before decoupling,provided recently, showed that
even in that case for some oscillation parameters kinetic equilibrium
may be strongly broken, especially in the resonant case (Dolgov \&
Villante, 2003).

 Since the 
oscillation rate is energy dependent $\Gamma \sim \delta m^2/E$ 
the low energy neutrinos start to oscillate first,
and later the
oscillations concern  more and more energetic neutrinos. 
 Hence,
the neutrino energy spectrum $n_{\nu}(E)$ may strongly deviate 
from its equilibrium form (see Fig.1), in case oscillations proceed between  
nonequilibrium neutrino states. 
\ \\
\ \\
\hspace{-0.8cm}\includegraphics[scale=0.25]{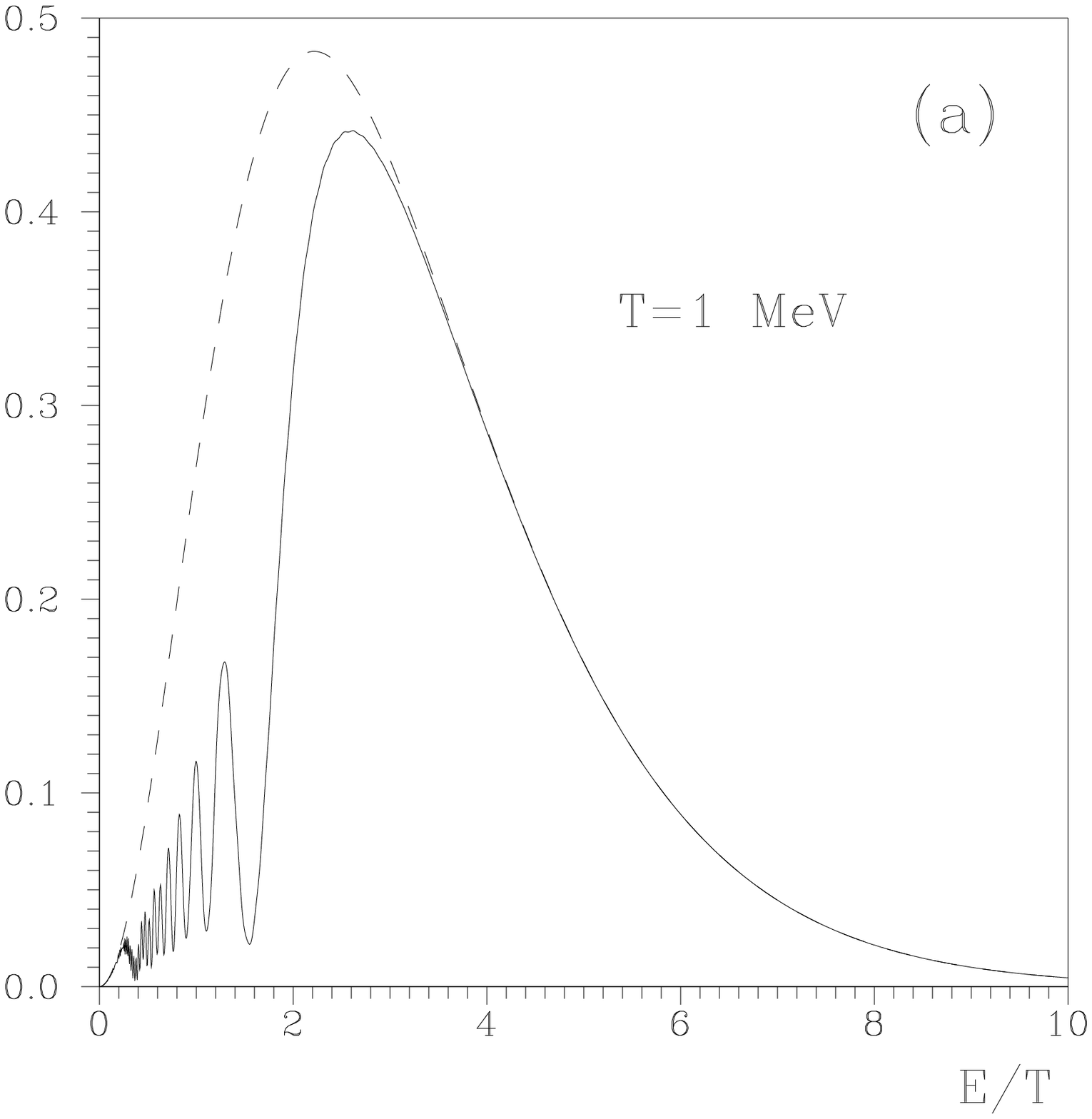}
\hspace{0.5cm}\includegraphics[scale=0.25]{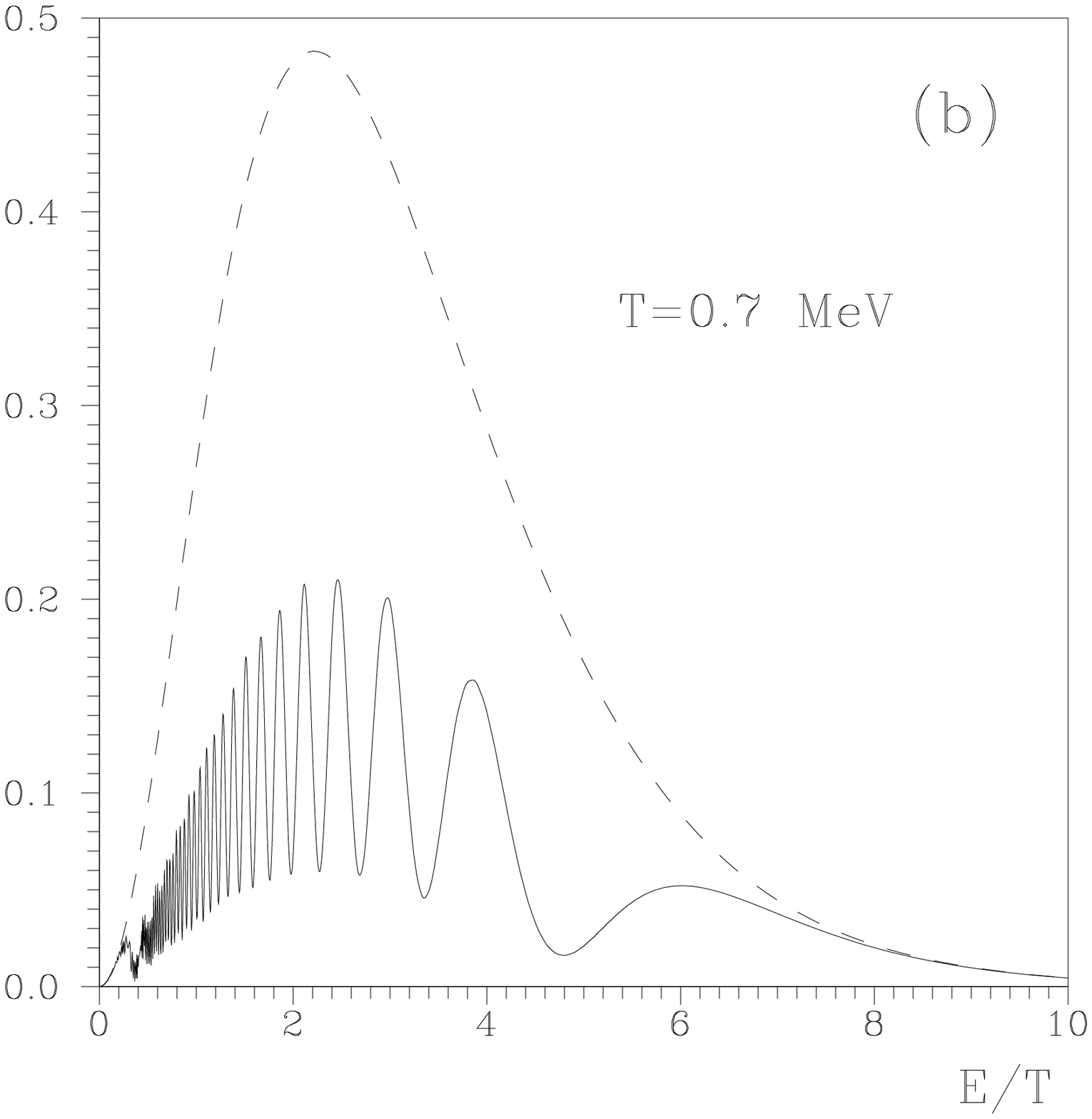}
\hspace{0.5cm}\includegraphics[scale=0.25]{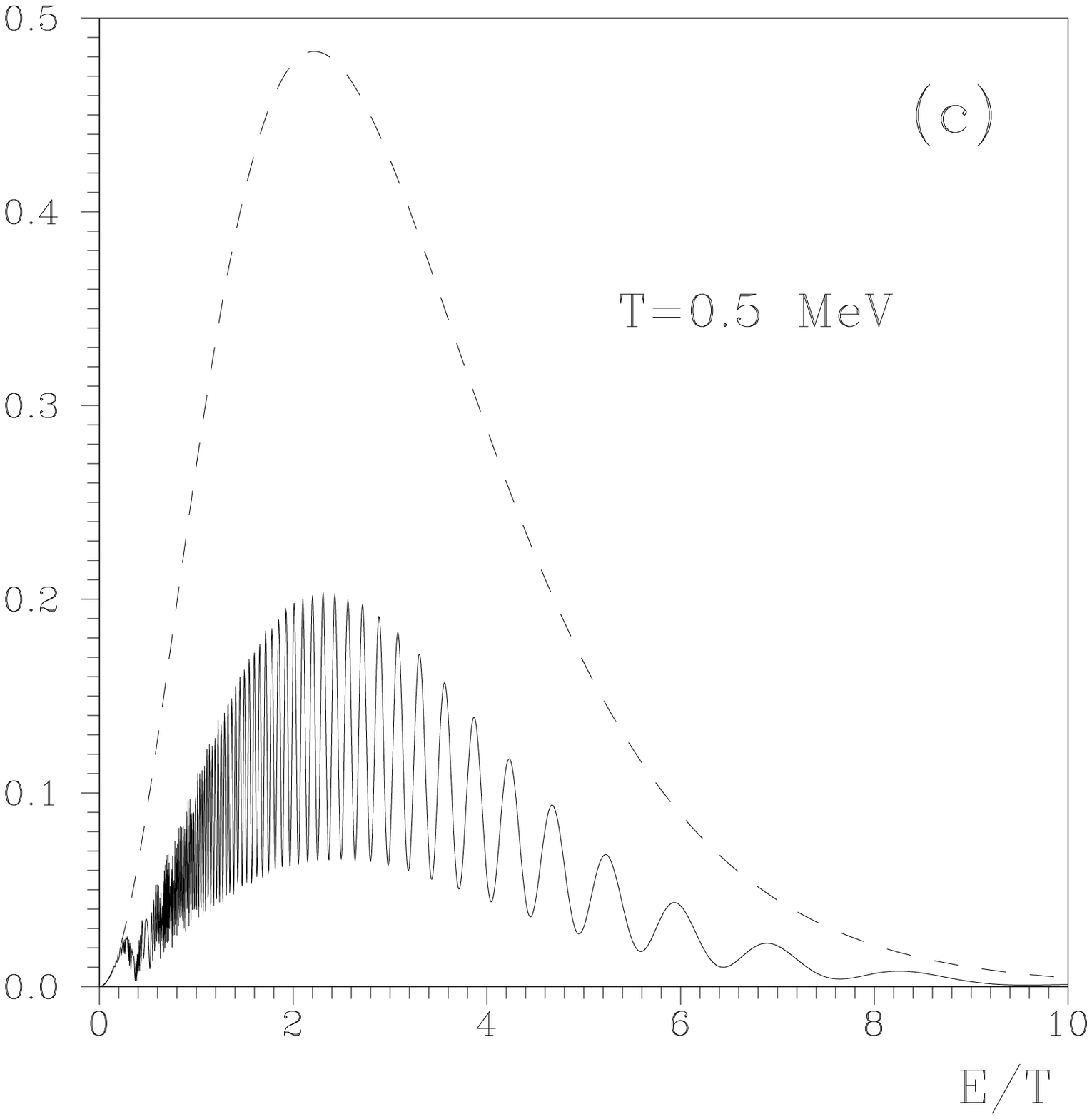}
\ \\
{\footnotesize Fig.1. The figures illustrate the degree of distortion
of the electron neutrino energy spectrum $x^2\rho_{LL}(x)$, where
$x=E/T$,
caused by oscillations with mass difference
$|\delta m^2|=10^{-7}$ eV$^2$ and mixing $\sin^22\vartheta=0.1$
for $\delta N_s=0$. The
evolution of the spectrum through the period of nucleons freezing
is presented at characteristic temperatures $1$, $0.7$ and $0.5$
MeV.
The dashed curve gives the
equilibrium spectrum at the given temperature.}
\ \\

The   distortion leads both to a {\it depletion of the active neutrino
number densities} $N_{\nu}$:
$$
N_{\nu}\sim \int {\rm d}E E^2 n_{\nu}(E)
$$
and a decrease of the $\Gamma_w$.
Thus spectrum distortion
 influences the nucleons kinetics, causing  an earlier
$n/p$-freezing and an overproduction  of
 $^4\!$He  yield. 

The spectrum distortion may also cause  
underproduction of He-4, when due to oscillations the
energy of the greater part of the neutrinos becomes smaller than the 
threshold  for the reaction
$\tilde{\nu}_e+p \rightarrow n+ e^+$ and the $n/p_f$-ratio decreases 
leading
to a decrease of He-4. However this effect is a minor one. Hence, the
 total effect is an overproduction of He-4.

The spectrum distortion is the greatest, in
 case the sterile state is empty at the start of oscillations,
$\delta N_s=0$.  It  decreases with the increase of the
degree of population of the sterile state at the onset of oscillations
(see Kirilova, 2002) as illustrated in  the following figures (Fig.2).
\ \\
\ \\
\hspace{-0.8cm}\includegraphics[scale=0.25]{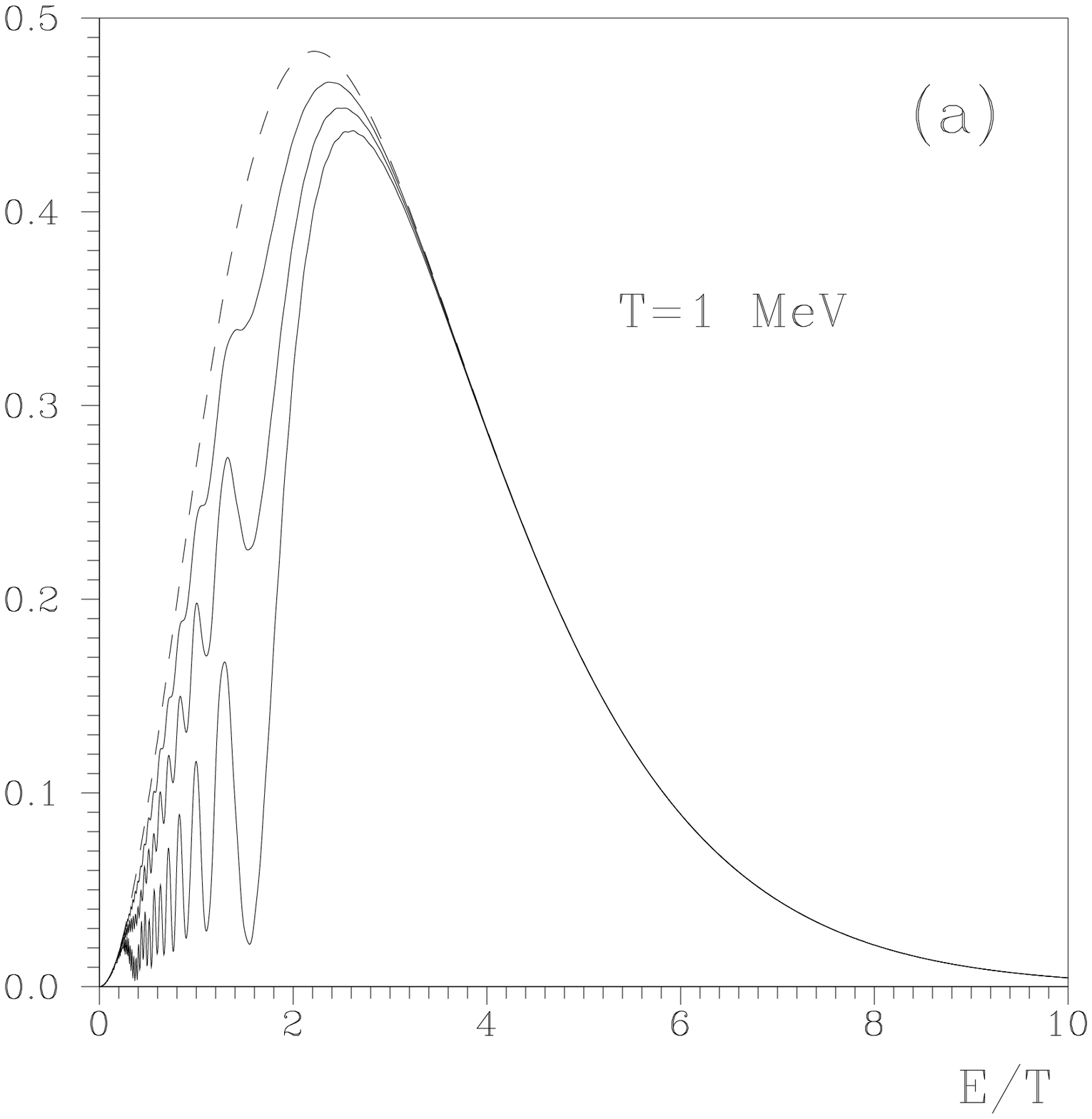}
\hspace{0.5cm}\includegraphics[scale=0.25]{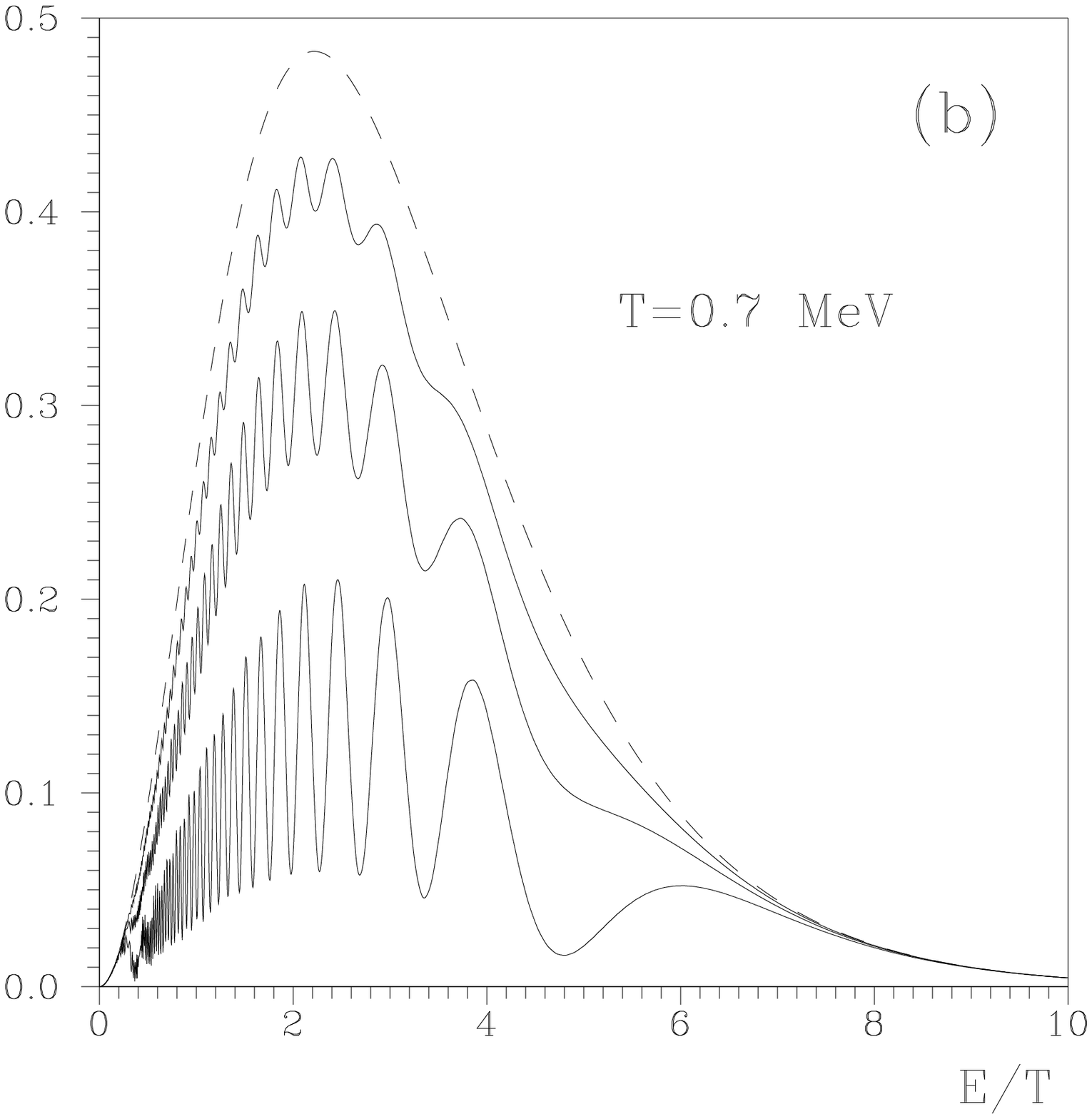}
\hspace{0.5cm}\includegraphics[scale=0.25]{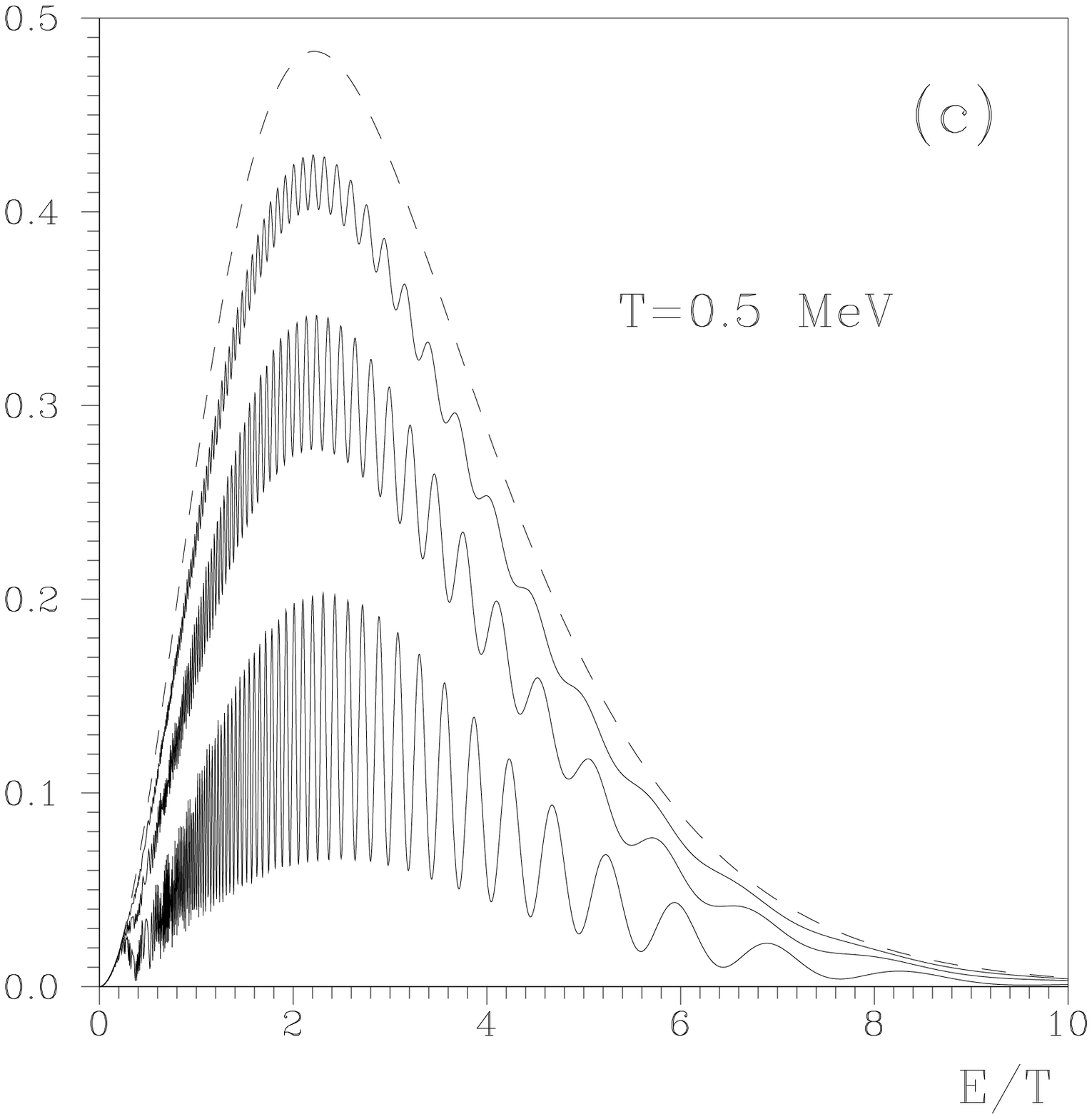}
\ \\
{\footnotesize Fig.2. The figures illustrate  the spectrum distortion at 
different
degrees of population of the steriles, namely $\delta N_s=0$
(lower curve), $\delta N_s=0.5$ and $\delta N_s=0.8$ (upper curve).
The dashed curve gives the equilibrium spectrum for comparison.
It is obvious that
the distortion of the spectrum is considerable and with time involves the
whole neutrino ensemble.}
\ \\
\ \\
 Spectrum distortion effect may be considerable both for the 
vacuum oscillations
(Kirilova, 1988)
 and oscillations in a medium 
 (Kirilova and Chizhov, 1996), both for the nonresonant (Kirilova \& 
Chizhov, 
1998) and resonant oscillations case (Kirilova \& Chizhov, 2000).
\ \\
{\it  (c)~Production of neutrino-antineutrino asymmetry}: 
Neutrino-antineutrino asymmetry may be generated
during the  resonant transfer of neutrinos (Miheev \& Smirnov, 1986;
Langacker et al., 1987, Kirilova \& Chizhov, 1996, Foot et al., 1996).
Dynamically produced asymmetry exerts back effect to
oscillating neutrino  and may change its
oscillation pattern. It influences $\nu$ and $\bar{\nu}$ number 
density evolution, their spectrum distorsion and the oscillation pattern 
-- all playing important role in $n-p$-kinetics.

Even when its value is not high enough to
have a direct kinetic effect on the synthesis of light elements, 
i.e. even when $L<<0.01$  it  
effects
indirectly BBN (Kirilova \& Chizhov, 1996, 1999, 2000). 
This dynamically produced asymmetry
suppresses
oscillations at small mixing angles,
leading to less overproduction of He-4 compared to the case without 
the account of asymmetry growth (see Fig.3), and hence aleviating BBN 
constraints on oscillation parameters. 

\ \\
\hspace{2cm}\includegraphics[scale=0.45]{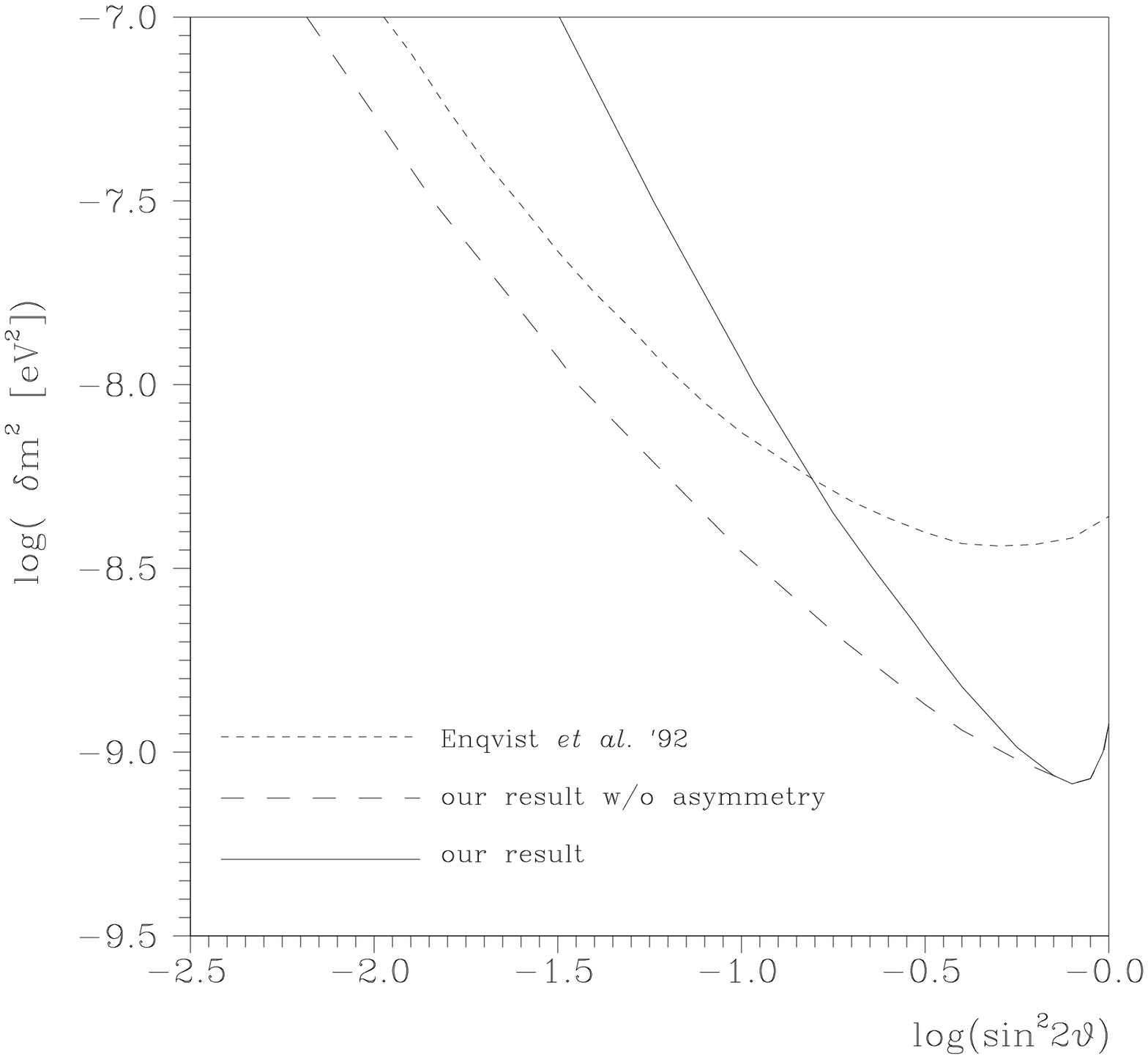}
\ \\
{\footnotesize Fig.3. On the $\delta m^2$ -- $\vartheta$ plane isohelium 
contour 
$Y_p=0.24$ is plotted. The long dashed curve presents the same $Y_p$ without 
the account of the asymmetry growth, while the small dashed curve presents the 
results of previous study, where both the spectrum distortion and the
asymmetry 
growth were ignored.} 
\ \\

In case, however, of initial asymmetry slightly higher than the baryon
one ($L \sim 10^{-6}$),  $L$ can  also  enhance  oscillations
transfers, leading to an increased overproduction of He-4.
However, for naturally small initial $L \sim $ baryon asymmetry,   the 
asymmetry effect is 
a subdominant one.

So, spectrum distortion effect  is the
dominant one and for a wide range of oscillation parameters
it is considerable
during the period of  nucleons freezing and hence  effects
primordial nucleosynthesis.
 The rough calculations not accounting for spectrum distortion
effects  may
underestimate oscillations effect  on BBN even
by several orders of magnitude of $\delta m^2$ (Chizhov \& Kirilova,
1999).

\subsection{Production of He-4 in the presence of neutrino oscillations}

The case of oscillations effective before electron neutrino freezing, was 
considered both analytically 
(Barbieri \&Dolgov, 1990; Barbieri \&Dolgov, 1991)
 and   numerically (Enqvist et al., 1992) 
accounting presicely for a) and partially for b) (namely, 
estimating the depletion of the neutrino number 
densities, assuming equilibrium neutrino energy spectrum). 

Analytical description was found in the case of very small mixing angles 
and 'large' mass differences  $\delta m^2> 10^{-6}$ eV$^2$, and for the 
case without spectrum distortion effects  (Dolgov, 2002, 2003; Dolgov \& 
Villante, 2003). 

The  nonequilibrium picture of neutrino oscillation effects (a)-(c)
is hard to describe analytically.
For nonequilibrium neutrino oscillations effective after active neutrino 
decoupliung,  i.e. for $(\delta m^2/eV^2)sin^42\theta<10^{-7}$,  
  the spectrum distortion effect was shown to 
play a considerable role.   
For that case  a complete  selfconsistent
 numerical analysis of the kinetics of the oscillating
 neutrinos, 
the  nucleons freeze-out
and the asymmetry evolution 
was provided  (Kirilova \& 
Chizhov, 96, 98, 2000).  Kinetic equations for neutrino density
matrix and neutron number densities in {\it momentum space}
 (Kirilova \& Chizhov, 
96,97) were used  to make a proper
precise  account  for
spectrum distortion effect, neutrino depletion and
neutrino asymmetry at each neutrino momentum.

The production of the primordial $^4\!$He,
$Y_p$ in the presence of
$\nu_e \leftrightarrow \nu_s$
 oscillations, effective after $\nu_e$ decoupling,  was calculated.
The numerical analysis was provided
for the  temperature interval [$0.3$ MeV, $2$ MeV]. 
 He-4 production was calculated 
both in the 
nonresonant 
(Kirilova \& Chizhov, 1998) and
resonant (Kirilova and Chizhov, 2000) 
oscillation cases (see also Kirilova, 2003).

Primordial helium is  considerably overproduced  in the presence of
active-sterile neutrino oscillations due to the effects (a)-(c).
The kinetic
effect of neutrino oscillations $\delta N_{kin}$ due to spectrum 
distortion usually  comprises a major
portion of
the total effect, i.e. it plays the dominant role in the overproduction 
of   $^4\!$He.
  It can
 be larger than the one
corresponding to an additional degree of
freedom.
The overproduction is maximal for the case of initially empty
 $\nu_s$ state $\delta N_s=0$ (Kirilova, 2003).

In the {\it nonresonant case} 
the effect of oscillations is proportional
to the  oscillation parameters. It  becomes  very
small (less than $1\%$) for small mixings: as small as
$\sin^22\theta=0.1$ for $\delta m^2=10^{-7}eV^2$,
and  for small mass differences: $\delta m^2 <10^{-10}$ eV$^2$ at
maximal mixing. 
The effect is maximal at maximal mixing for a given mass 
difference.

 $Y_p^{max}$ increase with $\delta m^2$  till 
$\delta m^2=10^{-7}$ eV$^2$
in
our model. Further increase of the mass differences requires a decrease
of the
maximal mixing angle considered, such that oscillations remain effective
after $\nu_e$ decoupling  
($\sin^4 2\vartheta\le 10^{-7} \delta m^{-2}$).
 Therefore, for higher $\delta m^2$ in the discussed oscillation model
 $\vartheta<\pi/4$, and  $Y_p^{max}$ decreases with further increase of
 $\delta m^2$  beyond  $\delta m^2 \sim 10^{-7} $eV$^2$ (see Fig.4).

In the {\it resonant oscillation case}, however,  for a 
given $\delta m^2$
there exists some resonant mixing angle, 
at which the oscillations
effects are enhanced by the medium  due to the MSW effect (see 
Wolfenstein 1978; Mikheev \& Smirnov, 1985), and hence, 
the overproduction of
He-4 is greater than that corresponding to the vacuum 
maximal mixing angle. In case $N_s=0$   $\delta N_{kin}^{max}>1$ for
 $\delta m^2>10^{-9}$ eV$^2$.

He-4 overproduction in the resonant case  can be up to $31.8\%$, 
while   
 in the nonresonant one - up to $13.8\%$ (see Fig.4).
So, the maximum overproduction of  $^4\!$He
corresponds to an increase of the neutrino effective degrees
of freedom  $\delta N_{kin}^{max}\sim 6$.
\ \\
\ \\
\includegraphics[scale=0.35]{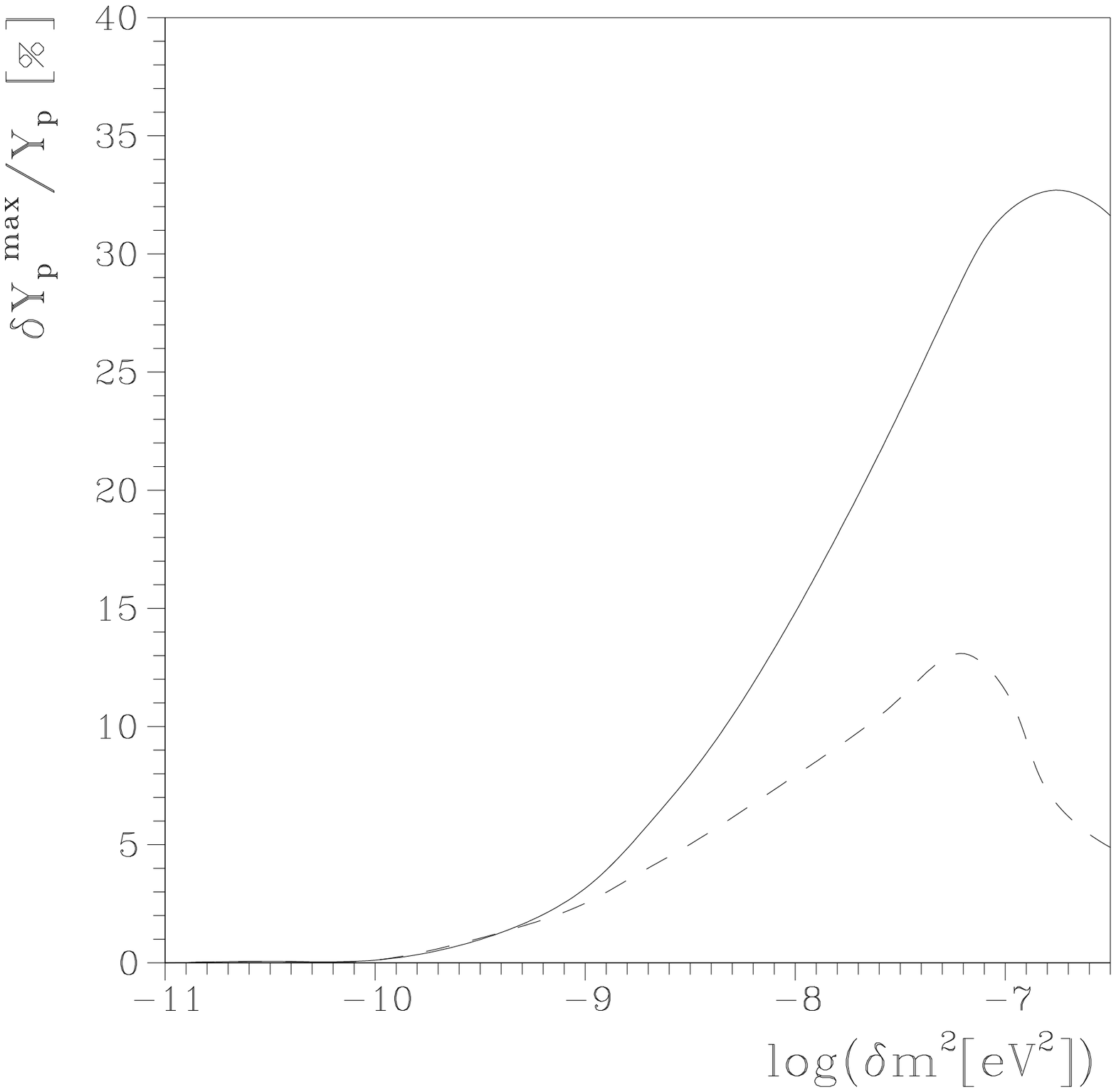}
\includegraphics[scale=0.35]{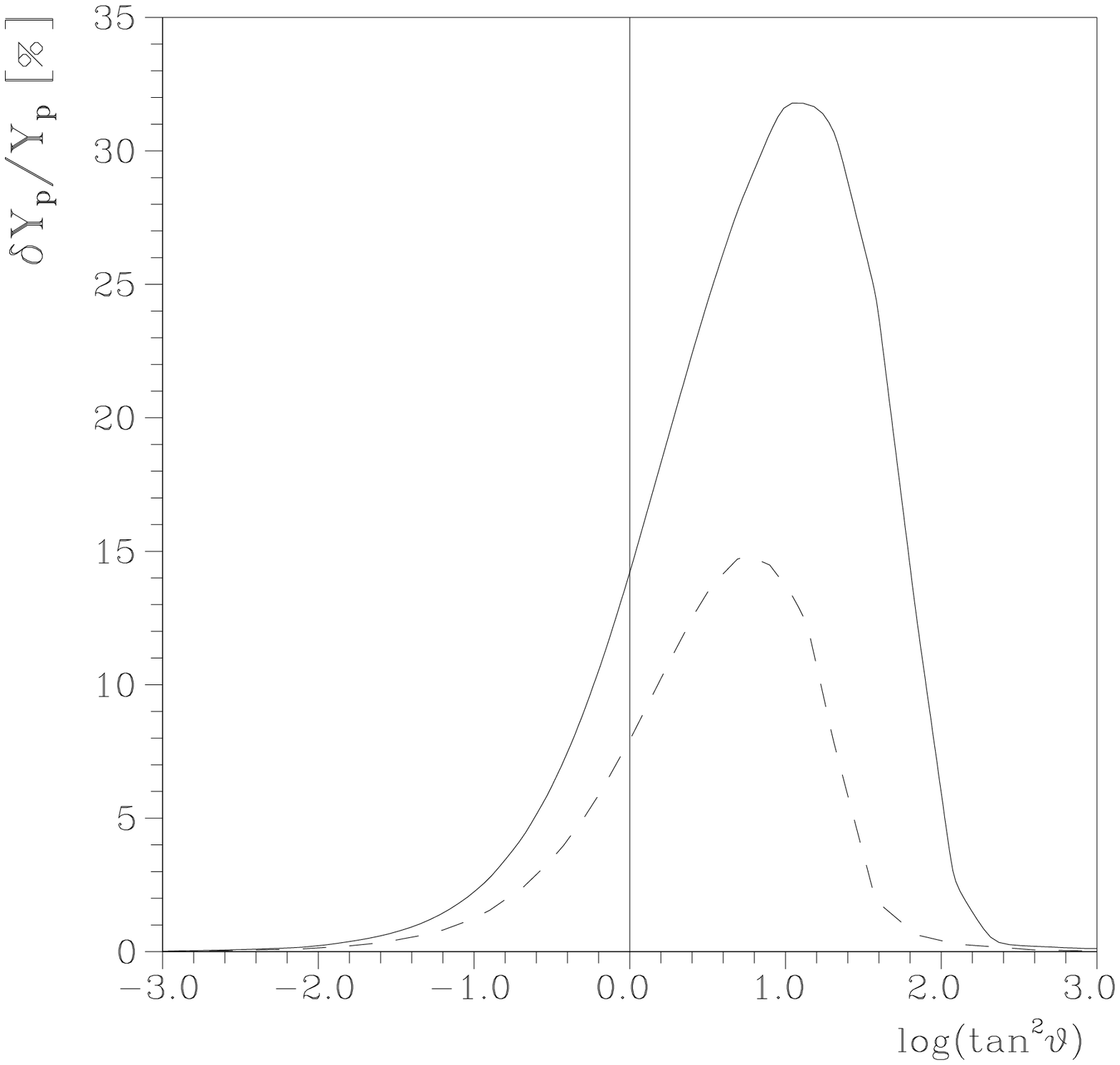}
\ \\
{\footnotesize Fig.4. In the l.h.s. figure  the
maximal  relative increase in the primordial $^4\!$He
as a function of neutrino mass differences:
$\delta Y_p^{max}/Y_p=\delta Y^{osc}_p/Y_p(\delta
m^2)$ is presented  for $\vartheta=\pi/4$ in the
nonresonant case, and for the resonant mixing angles in the resonant
case (upper curve). In  the r.h.s. figure maximum primordial  $^4\!$He 
abundance
for the resonant
and the non-resonant oscillation case, as a function
of the neutrino mixing angle at $\delta m^2=10^{-8}$ eV$^2$ (lower 
curve)  and $\delta m^2=10^{-7}$ eV$^2$
is given.}
\ \\
\subsection{COSMOLOGICAL CONSTRAINTS ON OSCILLATION PARAMETERS}  
\ \\

{\it Cosmological Constraints --- $\delta N_s=0$ Case}
\ \\

Observational  data on primordial He-4 put stringent limits 
 on the allowed active-sterile oscillation parameters.
First BBN limits were derived in the pioneer works (Barbieri \& Dolgov, 
1990, 1991; Enqvist et al., 1990, 1992) under the assumption of 
kinetic equilibrium (neutrinos were described by a single momentum state 
with the thermal average energy). 
These constraints  accounted for the effects 
a) and partially for b). 

They were recently updated (Dolgov, 2002; Dolgov \& Villante, 
2003). In the nonresonant case they can be approxilmated: 

$$
(\delta m^2_{\nu_e\nu_s}/eV^2)sin^42\theta^{\nu_e\nu_s}=3.16 . 10^{-5} 
(\delta N_{\nu})^2
$$
$$
(\delta m^2_{\nu_{\mu}\nu_s}/eV^2)sin^42\theta^{\nu_{\mu}\nu_s}=1.74 . 
10^{-5} (\delta N_{\nu})^2
$$

assuming kinetic equilibrium and using stationary point approximation. 
 The bounds  are reasonably accurate  for large mass differences in 
case of efficient repopulation of active neutrinos. For the exact 
constraints in the resonant case see (Dolgov \& Villante,
2003).  

However, as discussed in previous subsection, at lower mass differences, 
when 
sterile neutrino production takes place after active neutrino 
freezing, the re-population of active neutrino becomes slow and 
hence, kinetic 
equilibrium may be strongly broken due to active-sterile oscillations.  
The spectrum distorsion of $\nu_e$ may be considerable, leading to 
strong 
influence on nucleons kinetics (effects ( b) and (c)).
 The analysis of such oscillations with precise kinetic 
 accounting for the effects ( b) and (c) allowed to put  stringent 
constraints to oscillations parameters at small mass differences. 
We will discuss below these BBN constraints (Kirilova \& Chizhov, 1998, 
2000, 2001; Kirilova, 2003).

We assume  the uncertainty of observational helium-4 to be  
$\delta Y_p/Y_p^s<3\%$ in accordance with observations of helium and 
also with the resent WMAP constraints on the additional relativistic 
degrees of freedom (Crotty et al., 2003, Cyburt et al., 2003, Cuoco et 
al., 2003).
 Then    the
range of cosmologically excluded electron-sterile oscillations parameters  
is  situated above the 
$3\%$ contour at Fig.5:

\ \\
\hspace{1.6cm}\includegraphics[scale=0.55]{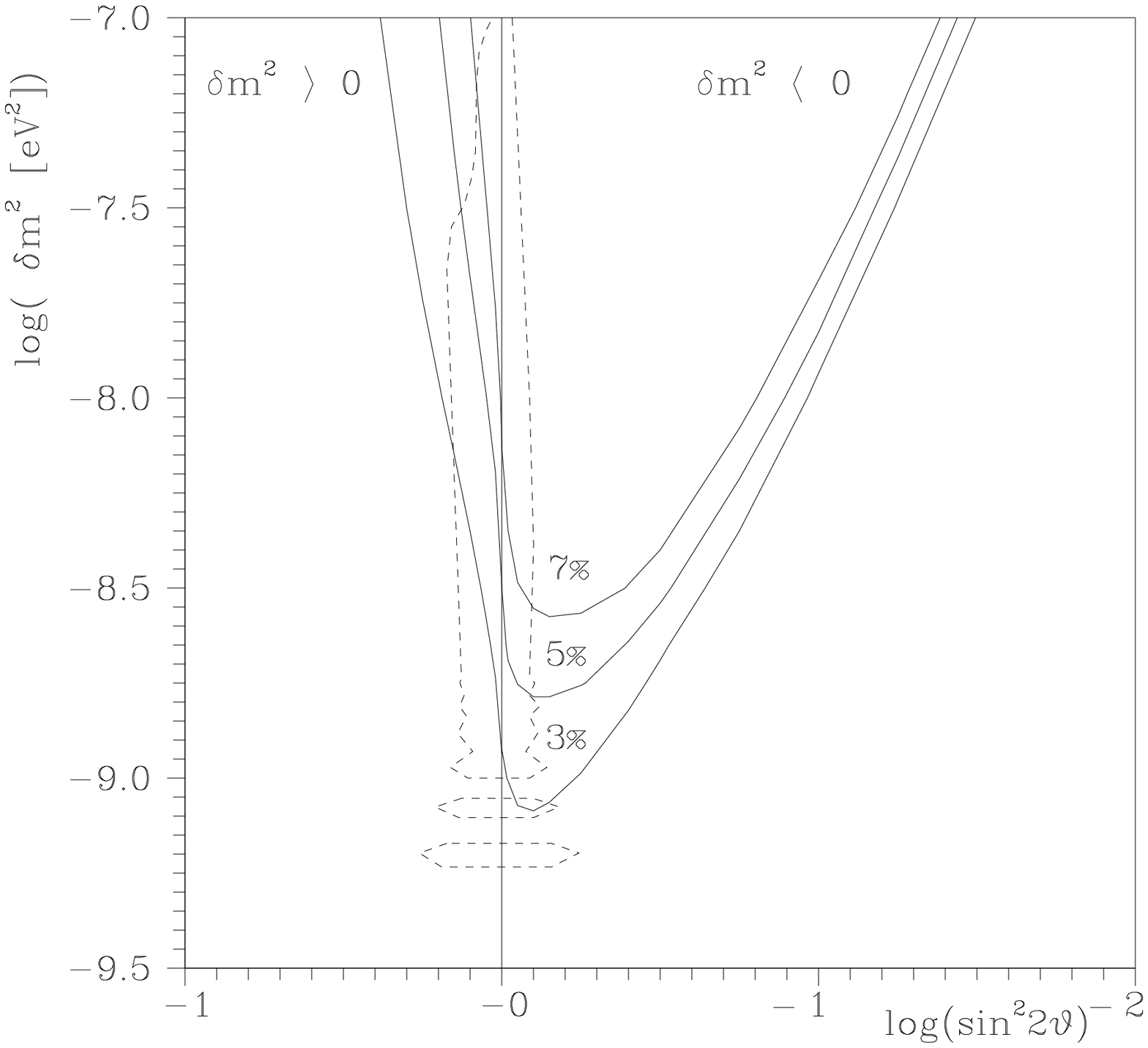}
\ \\
{\footnotesize Fig.5. The combined iso-helium contours  for the 
nonresonant and the resonant
case, for  $\delta Y_p=(Y_{osc}-Y_p)/Y_p=3\%, 5\%, 7\%$
(Kirilova \& Chizhov, 2001). The dashed curves present LOW sterile solution.}
\ \\

The analytical fits to the exact constraints are:
\begin{eqnarray*}
\delta m^2 (\sin^22\vartheta)^4\le 1.5 \times10^{-9} {\rm eV}^2
   ~~~\delta m^2>0  \\
                 |\delta m^2| < 8.2\times 10^{-10} {\rm eV}^2
               ~~~~\delta m^2<0,~~ large~~ \vartheta,
\end{eqnarray*}

Due to the precise account of the kinetic effects of oscillations, these  
constraints
are nearly  an order of magnitude stronger at large mixings than other 
numerical calculations (Enqvist et al., 1992) 
of 2-neutrino mixing  
and much 
constraining the mass differences values than the 
constraints derived  for oscillations effective before the
electron neutrino freeze-out (Dolgov, 2002; Dolgov  \& Villante 
2003).
 In the resonant
case,  due to the proper 
account
of the asymmetry  generated in oscillations, they are less restrictive 
at small
mixings \footnote{Mind that the oscillations generated asymmetry 
suppresses oscillations at small mixing angles, reflecting in less 
overproduction of He-4 in comparison with the case where asymmetry 
generation was neglected} than 
the  constraints (Enqvist 
et al., 1992).

The cosmological constraints
 exclude almost completely sterile LOW solution to the solar
neutrino problem, besides
the sterile LMA solution and sterile atmospheric
solution, excluded in previous
works.
  This result is consistent with the
global  analysis (Holanda \&  Smirnov, 2003; 
Giunti  \& Laveder, 2003; Maltoni et al., 2003; Bahcall et al., 2003; see 
also Fogli et al., 2001, 2003)
of the
 data from neutrino oscillations experiments KamLAND, SNO, 
SuperKamiokande, 
GALLEX+GNO, SAGE and Chlorine, which do not favour $\nu_e \leftrightarrow \nu_s$
 solutions.

These cosmological constraints should be generalized for the case of 
4-neutrino mixing\footnote{as was already done for the case of fast 
oscillations 
proceeding before neutrino decoupling in (Dolgov  \& Villante 
2003)}, since the effect of mixing between active neutrinos on 
the BBN constraints of electron-sterile oscillations has been proved  
 important in the resonant oscillation case (Dolgov  \& Villante
2003).
\ \\

{\it Cosmological Constraints  --- $\delta N_s\ne0$ Case}

 Sterile neutrinos
$\nu_s$
 may be present at the onset of
BBN epoch  --- they may be
 produced in GUT models,  in models with large extra
dimensions,  Manyfold Universe models,
 mirror matter models, or in
 $\nu_{\mu,\tau}\leftrightarrow \nu_s$ oscillations
 in  4-neutrino mixing
schemes.  Hence, the degree of population of $\nu_s$ may be different
depending on the $\nu_s$ production model.
Therefore, it is interesting to  study the  distortion of $\nu_e$
energy spectrum  due to oscillations  $\nu_e\leftrightarrow \nu_s$,
and its influence on BBN
 for different degree of population of the initially present sterile 
neutrinos $0 \le \delta N_s \le 1$.
$Y_p$,
 for different  $\delta N_s$ values and different
sets of oscillation parameters  $Y_p(\delta N_s, \delta m^2,
sin^22\vartheta)$ was calculated  (Kirilova, 2002). 

$\delta N_s \ne 0$ present before $\nu_{\mu,\tau}\leftrightarrow \nu_s$
just leads to an increase of the total energy density of the
Universe, and it is straightforward to re-scale the existing constraints.
In the  $\nu_e\leftrightarrow \nu_s$ oscillations case, however, the
presence of
$\nu_s$ at the onset of oscillations influences in addition
  the kinetic
effects of  $\nu_e\leftrightarrow \nu_s$ on BBN.
Larger  $\delta N_s$  decreases the kinetic
effects,
because the element of initial non-equilibrium between the active and the
sterile states is less expressed (see the dashed curves in Fig.6).
\ \\

\hspace{2cm}\includegraphics[scale=0.4]{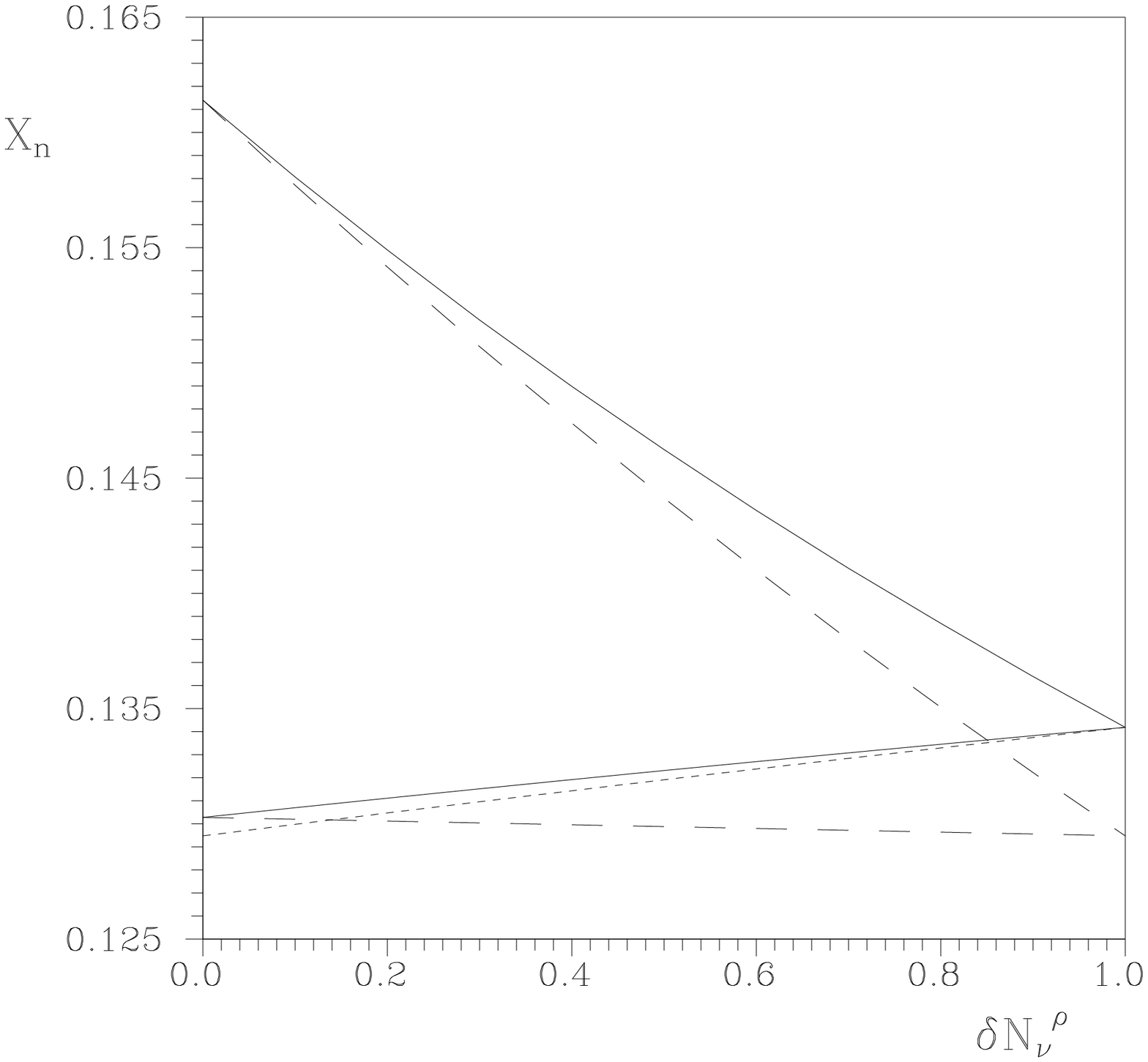}
\ \\
{\footnotesize Fig.6. The  solid curves present frozen
neutron number density relative to
nucleons $X_n^f=N_n^f/N_{nuc}$ as a function  of
the sterile neutrino initial population, at $\delta m=\pm 10^{-7}$
eV$^2$, $\sin^2 2\theta=10^{-1}$. The dashed curves present
the
kinetic effect, while the dotted curve presents
energy
density increase effect. The upper curves correspond to the
resonant case, the  lower ---to  the   non-resonant one.}
\ \\

{\it Neutrino spectrum distortion effect  is
very strong  even when there is a  considerable
population of the sterile neutrino state  before the
beginning of the electron--sterile oscillations.}
It always gives positive $\delta N_{kin}$, which for a large range of
initial sterile population values, is bigger than $1$.
  The kinetic
effects are
the  strongest for $\delta N_s=0$, they
 disappear for $\delta N_s=1$, when
$\nu_e$ and  $\nu_s$
states are in equilibrium, and the total effect reduces to the SBBN with
an additional neutrino (Fig.6).
\ \\

\hspace{2cm}\includegraphics[scale=0.4]{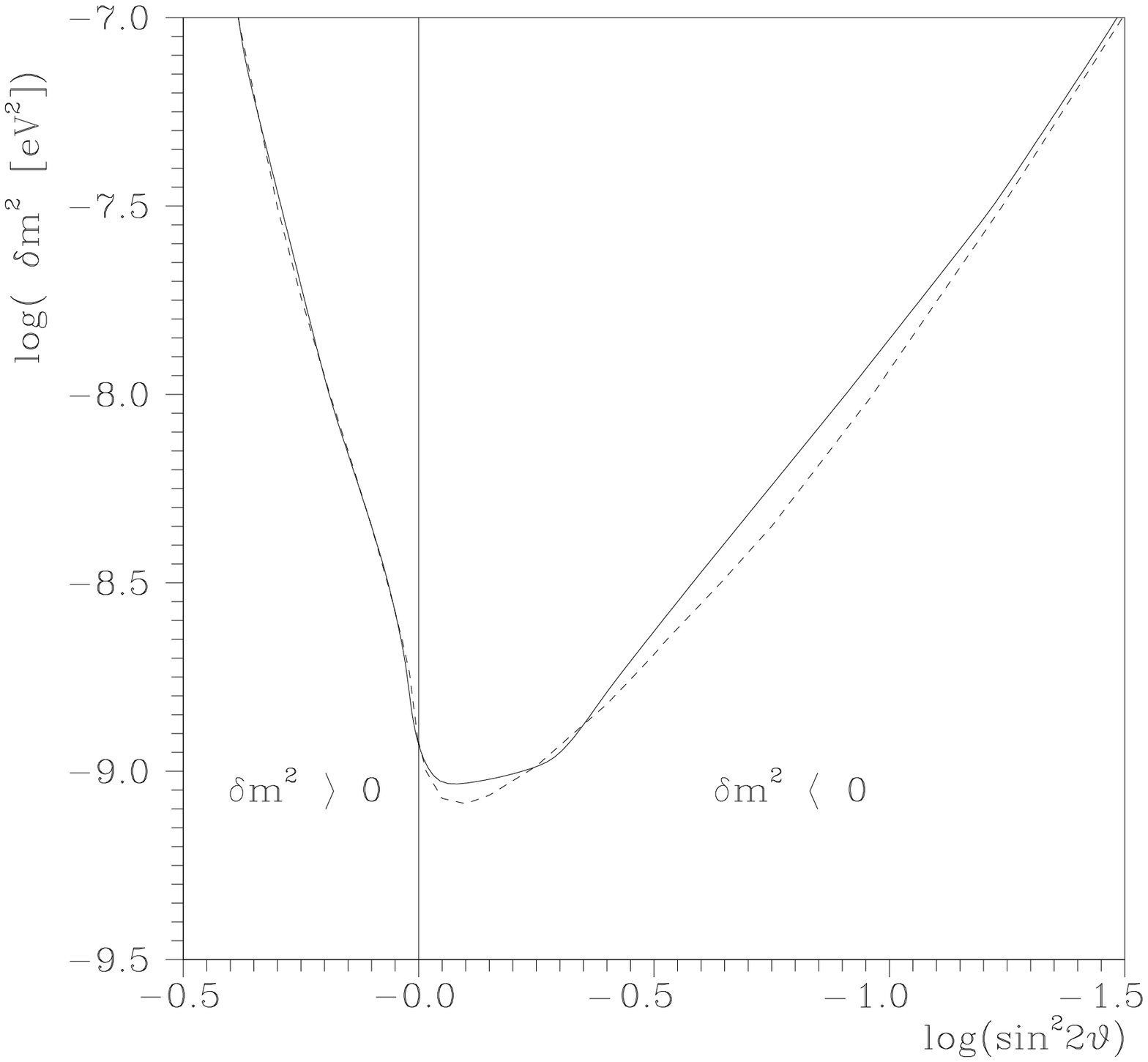}
\ \\
{\footnotesize Fig.7. The dashed contours present BBN constraints for  
$\delta N_s=0$,
the solid --- $\delta N_s=0.1$.}
\ \\

Our numerical analysis has shown that up to  $\delta N_s=0.5$
 the cosmological constraints are slightly
 changed and remain
stringent, as before. 

So, the presence of non-zero initial population of the setrile state 
does not change or remove the stringent cosmological constraints.

Cosmological constraints can be relaxed assuming  even higher than 
the 0.007  systematic error. However, even for $0.25$  
isohelium contour 
the constraints on LOW sterile solution are not removed, but just 
relaxed.  

Small relic fermion  asymmetry $L << 0.1$, however, is capable  
 to suppress oscillations with small mass differences,  thus removing 
BBN constraints for the corresponding oscillation parameters.  
For the oscillations effective after neutrino decoupling we have found 
that $L \sim 10^{-5}$ is large enough to suppress oscillations and 
remove BBN constraints (Kirilova \& Chizhov, 2001). 
So, we expect that  it is possible to remove 
  the cosmological constraints  only for rather small  $\delta m^2$, 
as far as asymmetry larger than $0.1$ 
is not allowed\footnote{Such asymmetry in $\nu_e$ sector changes 
unacceptably 
the kinetics of nucleons and the He-4 yield. While in $\nu_{\mu,\tau}$ 
sectors it is constrained to the same level on the basis of redistribution 
of the 
asymmetries due to oscillations (Dolgov et al., 2002)}.

So, the presence of non-zero initial population of the sterile state
does not change or remove the existing stringent cosmological constraints 
on active-sterile neutrino oscillations. They are not relaxed considerably 
also in case exagerated He-4 abundance is asumed. Even assuming 
unnaturally large lepton asymmetry ($B << L <0.1$), they are only relaxed, 
not removed. In case of naturally small L, BBN provides the most stringent 
constraints on  active-sterile neutrino oscillations parameters, and 
restricts the fraction of the sterile neutrinos in the neutrino anomalies. 

Besides, as discussed previously, cosmology provides still the most 
stringent upper limit for neutrino masses. So, it provides valuable 
precision probe of new neutrino physics.

\section{CONCLUSIONS}

During the last quarter of the 20th century strong evidence for non-zero 
neutrino masses and mixings has been provided from various 
neutrino oscillations 
experiments (astrophysical: solar neutrino experiments and atmospheric 
neutrino 
experiments and terrestrial: LSND and KamLAND). 

The essential role of 
neutrino oscillations for processes in the early Universe was realized. 
Stringent cosmological constraints on the oscillation parameters 
were obtained on the basis of BBN considerations. Cosmological analysis 
of CMB, LSS and BBN data  provide 
stringent upper limit for neutrino masses.
Non-zero neutrino masses and mixings, required by the experimental data 
and cosmology, 
 require new physics beyond the standard electroweak model.
   The scale of this NP 
is inversely proportional to the neutrino masses and appears to be 
of the order 
of the Grand Unification scale. 
Hence, the neutrino oscillation experiments and  cosmology  
 point the way towards  the Grand Unified Theory of elementary particles.

\section{ACKNOWLEDGEMENTS} 

It is my pleasure to thank  O. Atanackovi\'c-Vukmanovi\'c and 
the organizing committee of the NCYA Conference for the hospitality. I 
am grateful to I. Tkachev and the organizing committee of 
CAPP2003 for the financial support for my participation. 

D.K. is a regular associate of ICTP, Trieste, Italy.  
This work is supported in part by Belgian Federal Government (office for 
scientific affairs grant and IAP 5/27)   
\references

Aguilar A. et al. : 2001,  \journal{Phys. Rev. D} \vol{64}, 112007. 

Allen S., Schmidt R., Bridle S. : 2003, astro-ph/0306386. 

Ahmad Q. et al. : 2002, \journal{Phys. Rev. Lett.} \vol{89}, 011301. 

Ahn M. et al. : 2003,  \journal{Phys. Rev. Lett.} \vol{90}, 
041801. 

Bahcall J., Gonzalez-Garcia M., Pena-Garay C. : 2003, \journal{JHEP} 
\vol{0302}, 009.
 hep-ph/0212147

Balantekin A., Yuksel H. : 2003, hep-ph/0301072

Barbieri R., and Dolgov A.: 1990,  \journal{Phys. Lett.} \vol{B 237}, 440.

Barbieri R., Dolgov A. : 1991, \journal{Phys. Lett. B} \vol{237}, 440.

Barger V., Weiler T., Whisnant K. : 1998,  \journal{Phys. Lett. B} 
\vol{442}, 255. 
 
Barger V. et al.  : 2003, hep-ph/0305075.  

Bazarko A. : hep-ex/0210020.

Bilenky S., Giunti C., Grimus W.,  Schwetz T.: 1999, \journal{Astrop. Phys.} 
\vol{11}, 413.  

Bilenky S., Guinti C., Grifols J., Masso E. : 2003,  \journal{Phys. Rept.} 
\vol{379},  69.  

Bono G. et al. : 2002,  \journal{Ap. J.} \vol{568}, 463.

Cassisi S., Salaris M., Irwin A. : 2003, astro-ph/0301378.

Chizhov M.,  Kirilova D. : 1999,  hep-ph/9908525.

Coc A. et al. : 2003,  astro-ph/0309480. 

Cowsik R., McClleland J. : 1972 \journal{Ap. J.} \vol{180}, 7.

Crotty P., Lesgourgues J., Pastor S.: 2003,  \journal{Phys. Rev. D} 
\vol{67}, 123005.

Cuoco A. et al. : 2003, astro-ph/0307213.

Cyburt R., Fields B., Olive K. : 2003,  \journal{Phys. Lett. B} \vol{567}, 
227.

Di Bari P., 2002, \journal{Phys. Rev. D} \vol{65}, 043509. 

Dolgov A.D. :  1981, \journal{Sov. J. Nucl. Phys.} \vol{33}, 700.

Dolgov A. D. et al. : 1999, \journal{Nucl. Phys. B} \vol{548}, 385. 

Dolgov A.D. :  2001,  \journal{Phys. Lett. B} \vol{506}, 7.

Dolgov A. D. : 2002, \journal{Phys.Rept.} \vol{370}, 333; 
\journal{Surveys in High Energy Physics} \vol{17}, 91. 

Dolgov A. D. : 2003, hep-ph/0306154.

Dolgov A., Kirilova D. : 1988,  \journal{Int. J. Mod. Phys. A} \vol{3}, 
267. 

Dolgov A. et al.: 2002,  \journal{Nucl. Phys. B} \vol{632}, 363.

Dolgov A., Villante  F. : 2003, hep-ph/0308083.  

Eguchi K. et al. : 2003, \journal{Phys. Rev. Lett.} \vol{90}, 021802.

Elgaroy O. et al. : 2002, \journal{Phys. Rev. Lett.} \vol{89}, 061301.

Enqvist K., Kainulainen K., and Maalampi J.: 1990,  \journal{Phys.  Lett.}  
\vol{ B 244}, 186.

Enqvist K., Kainulainen K., Thompson M. : 1992,  \journal{Nucl. Phys. 
B}
\vol{373}, 498.

Esposito S. et al. : 2000, \journal{Nucl. Phys. B} \vol{568}, 421.

Fogli G., Lisi E., Marrone A. : 2001,  \journal{Phys. Rev. D} \vol{63}, 
053008.

Fogli  G. et al. : 2003,  \journal{Phys. Rev. D} \vol{67}, 073002.

Fogli G. et al. : 2003, hep-ph/0310012.

Foot R., Thompson M., Volkas R.: 1996, \journal{Phys. Rev. D} \vol{53}, R5349.

Fukuda Y. : 1998, \journal{Phys. Rev. Lett.} \vol{81}, 1562.

Fukugita M., Liu G.-C., Sugiyama N. : 2000, \journal{Phys. Rev. Lett.} 
\vol{84}, 1082.

Gamow G. : 1935, \journal{Ohio Journal of Science} \vol{35}, 406;\\
 \journal{Journal of the Washington Academy of Sciences} \vol{32}, 353, 
1942;\\
G. Gamow, \journal{Phys. Rev.} \vol{70}, 572, 1946.

Gerstein S., Zeldovich Ya. : 1966, \journal{JETP Letters} \vol{4}, 120.

Giunti C., Laveder M. : 2003,  hep-ph/0301276.

Giunti C. : 2003,   hep-ph/0309024.

 Gonzalez-Garcia M. C., Nir Y. : 2003, \journal{Rev. Mod. Phys.}  \vol{75},
345.

 Gyuk G.,  Turner M. : 1994 \journal{Phys. Rev.}  \vol{D 50}, 6130.  

Holanda P., Smirnov A. : 2003,  \journal{JCAP}  \vol{0302}, 001.

Holanda P., Smirnov A. : 2003a, hep-ph/0307266.  

Huey G., Cyburt R., Wandelt B. : 2003, astro-ph/0307080.

Izotov Yu. I., Thuan T. X. :  1998, \journal{Ap. J.} \vol{500}, 188.

 Kirilova D. : 1988,  JINR E2-88-301.

 Kirilova D., Chizhov M. : 1996, \journal{Neutrino96}, p.478; 1997; 
\journal{Phys. Lett. B} \vol{393}, 375;  CERN-TH-2002-209, hep-ph/0209104.

 Kirilova D., Chizhov M. : 1998, \journal{Phys. Rev. D} \vol{58}, 073004.

  Kirilova D.P., Chizhov M.V. : 2000, \journal{Nucl. Phys. B} \vol{591}, 457.

 Kirilova D.P., Chizhov M.V. : 2001, \journal{Nucl. Phys. B Suppl.} \vol{100},
 360.

 Kirilova D., Chizhov M. : 2001, Proc. Int. Conf. "Hot
 Points in Astrophysics", Dubna, Russia, p.56;  astro-ph/0108341.

 Kirilova D. : 2002,  hep-ph/0209104.

 Kirilova D. : 2003,  \journal{Astropart.Phys.} \vol{19}, 409.

 Kirkman D. et al. : 2003, astro-ph/0302006.

 Lisi E., Sarkar S.,  Villante F. : 1999,  \journal{Phys. Rev. D} 
\vol{59}, 123520.

 Lopez R. E., Turner M. S. : 1999, \journal{Phys. Rev. D} \vol{59}, 103502.

Maltony M., Schwetz T., Valle J.: 2003,  \journal{Phys. Rev. D} \vol{67}, 
093003.

 Mikheyev S., Smirnov A. : 1985, \journal{Sov. J. Nucl. Phys.} \vol{42},
913.


 Olive K., Skillman E., Steigman G. : 1997,  \journal{Ap. J.} \vol{483}, 
788.

 Pontecorvo B. : 1958, \journal{Sov. Phys. JETP} \vol{6}, 431.

 Prakash M. et al. : 2001, \journal{Ann. Rev. Nucl. Part. Sci.} 
\vol{51},  295.

 Raffelt G. : 2002  \journal{New Astron. Rev.} \vol{46}, 699.

Sarkar S. : 2003,  hep-ph/0302175. 

Shvartsman V. : 1969,  \journal{Pisma Zh.Eksp.Teor.Fiz.}  \vol{9}, 315;
 \journal{JETP Lett.}  \vol{9}, 184.

 Smirnov A. : 2003, hep-ph/0306075 

 Spergel D., et al., : 2003, astro-ph/0302209.

 Steigman G. et al. : 1986,   \journal{Phys. Lett. B} \vol{176}, 33. 

 Steigman G. : 2003,  astro-ph/0308511.

Terasawa N., Sato K. : 1987,  \journal{Phys. Lett. B} \vol{185}, 412.

 Trotta R., Hansen S. :  2003, astro-ph/0306588. 

 Wolfenstein L. : 1978, \journal{Phys. Rev. D} \vol{2369};

\endreferences
\end{document}